\documentclass[aps,prd]{revtex4}
\usepackage{epsfig,amsmath,amsfonts}
\newcommand{\fr}[2]{{\frac{#1}{ #2}}}
\newcommand{\Ref}[1]{(\ref{#1})}
\newcommand{\be}{\begin{equation}}
\newcommand{\ee}{\end{equation}}
\newcommand{\bn}{\begin{eqnarray}}
\newcommand{\en}{\end{eqnarray}}
\newcommand{\bd}{\begin{displaymath}}
\newcommand{\ed}{\end{displaymath}}
\newcommand{\bnn}{\begin{eqnarray*}}
\newcommand{\enn}{\end{eqnarray*}}
\newcommand{\bs}{\begin{subequations}}
\newcommand{\es}{\end{subequations}}
\newcommand{\adb}{\allowdisplaybreaks }
\begin{document}
\title{Dynamic Tsallis entropy for simple model systems}
\author{Nail R. Khusnutdinov}\email{nail@kazan-spu.ru}
\author{Renat M. Yulmetyev}\email{rmy@kazan-spu.ru}
\author{Natalya A. Emelyanova}\email{natali@kazan-spu.ru}
\affiliation{Department of Theoretical Physics, Kazan State
Pedagogical University, Mezhlauk 1, 420021, Kazan, Russia}
\begin{abstract}
In this paper we consider the dynamic Tsallis entropy and employ
it for four model systems: (i) the motion of Brownian oscillator,
(ii) the motion of Brownian oscillator with noise, (iii) the
fluctuation of particle density in hydrodynamics limit as well as
in (iv) ideal gas. We show that the small value of parameter
nonextensivity $0<q<1$ works as non-linear magnifier for small
values of the entropy. The frequency spectra become more sharp and
it is possible to extract useful information in the case of noise.
We show that the ideal gas remains non-Markovian for arbitrary
values of $q$.
\end{abstract}

\maketitle
\section{Introduction}
There is a great interest towards to studying complex systems of
physical, chemical, biological, physiological and financial origin
by using different approaches and concepts \cite{BecSch93,Kau93}.
Significant role in these investigations plays concept of
information entropy such as Shannon and Renyi entropy,
Kolmogorov-Sinai entropy rate
\cite{ScaSeaFer72,Kli98,Tsa99,GoyHan00,LatBarRapTsa00}. In the
Refs. \cite{YulKle98,YulGaf99,YulGafYulEme02} the notion of
dynamical information Shannon entropy \footnote{In literature
there is standard notion of dynamic entropy (see e.g. Ref.
\cite{KatHas95}). We suggest and use the functions $S$ and $S_q$
which we call the dynamic Shannon entropy and dynamic Tsallis
entropy.} has been defined for complex systems. It was suggested
to generalize the Shannon entropy by considering square of time
correlation function (TCF) as a probability of state. Then the
entropy becomes a function of time. The dynamic Shannon entropy
has successfully used to obtain new information in dynamics of
$RR$-intervals from human ECG, physiological activity of the
individuals and short-time human memory
\cite{YulKle98,YulGaf99,YulGafYulEme02}. The investigations reveal
also the great role of frequency spectrum of dynamic Shannon
entropy.

A new approach for entropy was suggested by Tsallis in Ref.
\cite{Tsa88} (the earlier works on this subject see in Ref.
\cite{AczDar75}). This entropy is characterized by one parameter
$q$ which was called as nonextensive parameter. The natural
boundary of this parameter is unit. It means that for $q\to 1$ the
Tsallis entropy transforms to Shannon entropy. Nevertheless, it
was often considered the cases with $q>0$ and even the limit
$q\to\infty$. There is/was huge activity for application of this
new notion to various complex systems (see, for examples, Ref.
\cite{Tsa99} and reference therein). It was shown that this
entropy gives a new kind of distribution. Many complex systems in
nature are described by this non-Gaussian distribution with
different values of parameter $q$.

The goal of this paper is generalization of dynamic Shannon
entropy in the manner as Tsallis entropy generalizes the Shannon
entropy. In other words we consider the Tsallis entropy with
square of TCF as probability of state. We refer this entropy as
dynamic Tsallis entropy (DTE). The real systems in nature have
very complicate behavior -- the useful signal is lost in "see"\/
of noises and random unexpected influences. For this reason we
would like to consider the dynamic Tsallis entropy for model
systems, which nevertheless have deep physical contents.
Application of the present theory to the alive systems was
considered in Ref. \cite{YulEmeGaf04}. We employ the information
Tsallis entropy and its frequency spectrum with different values
of parameter $q$ for model systems. We exploit four well-known
models. First one is the TCF of position of oscillator which
performs the Brownian motion. In order to consider more real
situation we use the motion of Brownian oscillator with noise in
second model. The noise is modelled by generator of random
numbers. Third model is the TCF of relaxation of particle density
fluctuations in hydrodynamics limit (the Landau-Placzek formula).
We consider a specific medium -- Helium. The fourth model
describes the relaxation of particle density fluctuations in ideal
gas. In the later case we additionally calculate the spectrum of
non-Markovity parameter for different $q$. The non-Markovity
parameter and its spectrum firstly was suggested in Refs.
\cite{ShuYul90,ShuYul91} and it was calculated for ideal gas in
Ref. \cite{YulKhu94}. This parameter characterizes the statistical
memory effects in systems. In present paper we use slightly
different approach and define the non-Markovity parameter in terms
of the entropy.

In all considered models the behavior of the dynamic Tsallis
entropies and their frequency spectrum crucially depend on the
parameter $q$. We observe that decreasing of this parameter $q\to
0$ allows us to enlarge the fine structure of entropies and to
make their frequency spectrum more sharp.

The organization of this paper is as follows. In section
\ref{Sec:2} we discuss the well known hierarchy of Zwanzig-Mori's
kinetic equations and general properties of the DTE and define
different kind of relaxation times. The TCF of Brownian motion of
oscillator is considered in section \ref{Sec:3} and with noise in
section \ref{Sec:4}. The formula of Landau and Placzek for
relaxation of density fluctuations are exploited in section
\ref{Sec:5}. The TCF of density fluctuations in ideal gas is used
in section \ref{Sec:6}. We finish our paper by concluding remarks
in last section \ref{Sec:7}.

\section{The Zwanzig-Mori hierarchy and Dynamic Tsallis Entropies}\label{Sec:2}

At the beginning we shortly discuss the well known hierarchy of
Zwanzig-Mori equations \cite{Zwa61,Mor65}. Let us consider the
dynamic variable $\delta A(t)$. It may be, for example, the
Fourier component of density fluctuation. This variable obeys to
the Liouville equation:
\bd
\fr{d\delta A(t)}{dt} = i\widehat{L} \delta A(t).
\ed
Applying $n$ times the Liouville operator $\widehat{L}$ to the
initial variable $\delta A(0)$ we obtain the infinite set of
variables $B_n(0)$
\bd
B_n(0) = \widehat{L}^n \delta A(0),
\ed
by using which and the Liouville equation we may obtain the
initial dynamical variable in arbitrary moment of time
\bd
\delta A(t) = \sum_{n=0}^\infty \fr{(it)^n}{n!}B_n.
\ed
Applying the Gram-Schmidt orthogonalization procedure
\cite{ReeSim72} to this set of functions we obtain the complete
set of dynamic variables $W_n$ which are orthogonal at the initial
time
\bd
\langle W_n^* W_l^{\phantom{*}}\rangle = \langle |W_n|^2 \rangle
\delta_{n,l},
\ed
where $\langle \ldots \rangle$ denotes the statistical average
over Gibbs ensemble. If the dynamic variable is evaluated by the
Liouville's operator, then the orthogonality is preserved at any
moment of time due to the self-adjointness of the Liouville's
operator.

The time correlation function $M_0$ of variable $W_0 = \delta A$
is defined as follows:
\bd
M_0(t) = \fr{\langle W_0(0)^* W_0(t) \rangle}{\langle |W_0(0)|^2
\rangle} = \fr{\langle W_0^* \exp{(i\widehat{L} t)}W_0
\rangle}{\langle |W_0|^2 \rangle}.
\ed
It is well known that this TCF obeys to the infinite hierarchy of
the Zwanzig-Mori's kinetic equations
\be
\frac{dM_n(t)}{dt} = i \omega_0^{(n)}M_{n}(t) - \Omega_{n+1}^2
\int_0^t d t' M_{n+1}(t') M_n (t-t'), \label{MainSystem}
\ee
where
\bd
\omega_0^{(n)} = \fr{\langle W_n^* \widehat{L} W_n
\rangle}{\langle |W_n|^2 \rangle},\ \Omega_n^2 = \fr{\langle
|W_n|^* \rangle}{\langle |W_{n-1}|^2 \rangle}
\ed
and
\be
M_n(t) = \fr{\langle W_n^* \exp{(i\widehat{L}_{22}^{(n)} t)}W_n
\rangle}{\langle |W_n|^2 \rangle}. \label{TCF}
\ee
The operator $\widehat{L}_{22}^{(n)}$ is defined in following
manner:
\bd
\widehat{L}_{22}^{(n)} = P_{n-1} P_{n-2} \ldots P_0 \widehat{L}
P_0 \ldots P_{n-2} P_{n-1}
\ed
in terms of projectors $P_n = 1 - \Pi_n$, where the $\Pi_n$ is
projector for state $W_n$:
\bd
\Pi_n = \fr{W_n\rangle \langle W_n^*}{\langle |W_n|^2 \rangle}.
\ed
The functions $M_n$ for $n\geq 1$ are not, in fact, usual TCFs
because the operator $\exp{(i\widehat{L}_{22}^{(n)}t)}$ is not
operator of evolution. Nevertheless, we will refer for these
functions as TCF's for the next dynamic variables.

The functions $M_n$ are considered as functions characterizing the
statistical memory of the system. In order to describe
quantitatively the non-Markovity of hierarchy, the parameter of
non-Markovity and its spectrum were introduced in Refs.
\cite{ShuYul90,ShuYul91}. The spectrum \footnote{This is not
frequency spectrum. Here the spectrum means set of parameters
$\epsilon_n$.} of this parameter is defined as ratio of two
neighboring relaxation times
\be
\epsilon_n = \fr{\tau_n}{\tau_{n+1}}.\label{Epsilon}
\ee
The relaxation time $\tau_n$ of function $M_n$ was defined as real
part of the Laplace image (see Eq.\Ref{Laplace}) of this function
at zero point:
\bd
\tau_n = \Re \widetilde{M}_n(0) = \Re \int_0^\infty M_n(t) dt,
\ed
where $\Re$ means a real part. Because of the function $M_{n+1}$
is an integral core of integro-differential equation
\Ref{MainSystem} for $M_{n}$, then this parameter \Ref{Epsilon}
compares the integral core with function itself. More precisely we
compare squares under TCFs. If parameter $\epsilon_n$ is around
unit, this level (level means $n$) is non-Markovian: we can not
transform integro-differential equation for $M_{n}$ to
differential one, and vice versa, if this parameter tends to
infinity the core has sharp peak and we may transform the
integro-differential equation for $M_{n}$ to differential one. In
this case there is no memory (integral) in this level. In this
paper we use slightly different definition for relaxation time
(see Sec.\ref{Sec:6}).

Applying the Laplace transformation
\be
\widetilde{M}_n(s) = \int_0^\infty dt e^{-st}M_n(t)\label{Laplace}
\ee
to equation \Ref{MainSystem} we transform this hierarchy to the
infinite system of algebraic equations
\be
\widetilde{M}_n(s) = [s-i \omega_0^{(n)} + \Omega_{n+1}^2
\widetilde{M}_{n+1}(s)]^{-1},\label{Laplac}
\ee
by using which we may express $M_n$ in term of zero TCF $M_0$.

In statistical physics of non-equilibrium systems the time
correlation  function (TCF) acts as the function of distribution
and pair correlation  and can be used to calculate different
thermodynamical parameters and the spacial structure of the system
\cite{Zwanzig65}. For many physical discrete systems it is
impossible to find a distribution function. For this reason it
seems optimal to obtain the TCF for investigations of complex
systems with the help of integro-differential equations which are
based on small increments of time and independent variables.

The set of the measured parameters of the complex system may be
represented as a set of fluctuations  \cite{YulEmeGaf04}
\be
Z=\{\zeta(T), \zeta(T+\tau), \zeta(T+2\tau), \cdots,
  \zeta(T+k\tau), \cdots, \zeta(T+\tau N-\tau)\},   \label{1.3.fluc}
\ee
where $\zeta$ represents the fluctuation $\delta x$ of some
quantity $x$.

Let us define the sampling by length  $k$,  which starts at the
moment $T+m\tau$ by the relation
\bd
\zeta_{m+k}^m=\{\zeta(T+m\tau), \cdots,  \zeta(T+m\tau
+(k-1)\tau)\}.
\ed
The operator which projects the sampling by length  $k$ on the
sampling at the initial moment of time has the form
\bd
\Pi=\frac{|\zeta_k^0\rangle\langle\zeta_k^0|}{\langle\zeta_k^0\zeta_k^0\rangle},
\ed
where angle brackets mean the scalar product (time average). The
utilization of this projection operator allows us to represent the
sampling as a sum of two independent parts
\bd
\zeta_{m+k}^m=\zeta_{m+k}^m{}'+\zeta_{m+k}^m{}'',
\ed
where
\bs
\bn
\zeta_{m+k}^m{}' &=& \Pi\zeta_{m+k}^m = \zeta_k^0 M_0(t),  \label{zeta'} \\
\zeta_{m+k}^m{}'' &=& (1-\Pi)\zeta_{m+k}^m = \zeta_{m+k}^m -
\zeta_k^0 M_0(t), \label{zeta''}
\en
\es
and TCF $M_0(t)$ is defined by \Ref{TCF}.

It is easy to see that
\bd
 \langle\left(\zeta_{m+k}^m\right)^2\rangle =
 \langle\left(\zeta_{m+k}^m{}'\right)^2\rangle
 + \langle\left(\zeta_{m+k}^m{}''\right)^2\rangle,
\ed
which is the consequence of orthogonality  of \Ref{zeta'} and
\Ref{zeta''}.

Direct calculations yield
\bd
\langle\left(\zeta_{m+k}^m{}'\right)^2\rangle =
\langle\left(\zeta_{k}^0\right)^2\rangle M_0(t)^2.
\ed
For stationarity processes,  when dispersion does not depend on
time,  we obtain
\bd
\langle\left(\zeta_{m+k}^m{}''\right)^2\rangle =
\langle\left(\zeta_{m+k}^m\right)^2\rangle -
\langle\left(\zeta_{k}^0\right)^2\rangle M_0(t)^2 =
\langle\left(\zeta_{m+k}^m\right)^2\rangle(1-M_0(t)^2).
\ed
Therefore the mean-square value of the fluctuations is presented
as a sum of two parts
\be
\langle\left(\zeta_{m+k}^m\right)^2\rangle =
\langle\left(\zeta_{k}^0\right)^2\rangle M_0(t)^2 +
\langle\left(\zeta_{k}^0\right)^2\rangle (1-M_0(t)^2),
\label{zet1}
\ee
or in a generalized form
\be
\langle\left(\zeta_{m+k}^m\right)^2\rangle =
\langle\left(\zeta_{k}^0\right)^2\rangle M_0(t)^2 +
\langle\left(\zeta_{m+k}^m\right)^2\rangle
(1-M_0(t)^2).\label{zeta^3}
\ee

The above expression \Ref{zeta^3} has a standard form of a mean
value
\be
\langle\left(\zeta_{m+k}^m\right)^2\rangle =
\langle\left(\zeta_{k}^0\right)^2\rangle P(t) +
\langle\left(\zeta_{m+k}^m\right)^2\rangle Q(t),  \label{prob}
\ee
where $P(t)$ is the probability of the state and $Q(t) = 1- P(t)$.
By analogy with this formula we define
\be
P_n(t)=|M_n(t)|^2\ (Q_n(t)=1-|M_n(t)|^2),\ n\geq0 \label{2a5}
\ee
as the probability of the creation (annihilation) of correlation
of fluctuations (or memory) for the $n$th level of relaxation
(see, for details Refs. \cite{YulKhu94}).

In the recent work of authors \cite{YulGafYulEme02} the dynamical
informational Shannon entropy for the study of complex systems was
suggested:
\bs \label{Shannon}
\be
S_n(t) = - \sum_{i= c,  a} P_i(t) \ln P_i (t) = - |M_n(t)|^2 \ln
|M_n(t)|^2 - \{1-|M_n(t)|^2\}\ln \{1-|M_n(t)|^2\},  \label{4}
\ee
where
\bn
S_n^c(t) &=& - |M_n(t)|^2 \ln |M_n(t)|^2,  \label{cc}\\
S_n^a(t) &=& - \{1-|M_n(t)|^2\}\ln \{1-|M_n(t)|^2\}. \label{ac}
\en
\es
Here the $S_n^c(t)$ is the entropy for the stochastic channels of
memory creation,  and $S_n^a(t)$ is entropy for the stochastic
channels of memory annihilation.

The infinite set of TCFs $M_n(t)$ produces the infinite set of
entropies which are defined by relations
\bd
S^n[t] = S[M_n(t)],\ S^{nc}[t] = S^c[M_n(t)],\ S^{na}[t] =
S^a[M_n(t)].
\ed
Because of the TCFs are always smaller then unit then all of these
entropies are positive.

We also generalize our definitions of entropy due to Tsallis by
introducing parameter of nonextensivity $q$:
\bs\label{QS}
\bn
S_{q}(t) &=&  -\fr{(1-|M_0(t)|^2)^q - 1 + |M_0(t)|^{2q}}{q-1},\adb \\
S^{c}_{q}(t) &=&  -\fr{|M_0(t)|^{2q}-|M_0(t)|^{2}}{q-1},\adb \\
S^{a}_{q}(t)&=&  -\fr{(1-|M_0(t)|^2)^q - 1 + |M_0(t)|^{2}}{q-1}.
\en
\es
The infinite set entropies are defined by relations
\bd
S^n_q[t] = S_q[M_n(t)],\ S^{nc}_q[t] = S^c_q[M_n(t)],\ S^{na}_q[t]
= S^a_q[M_n(t)].
\ed

We also calculate the frequency spectra $\widehat{S}^n[\nu],
\widehat{S}^{nc}[\nu], \widehat{S}^{na}[\nu],\widehat{S}^n_q[\nu],
\widehat{S}^{nc}_q[\nu], \widehat{S}^{na}_q[\nu]$ of these
functions and define the spectra of the relaxation times by
relations
\bnn
\tau_n &=& \widehat{S}^n[\nu]_{\nu=0},\  \tau_{nc} =
\widehat{S}^{nc}[\nu]_{\nu=0},\ \tau_{na} =
\widehat{S}^{na}[\nu]_{\nu=0},\adb\\
\tau_{qn} &=& \widehat{S}^n_q[\nu]_{\nu=0},\  \tau_{qnc} =
\widehat{S}^{nc}_q[\nu]_{\nu=0},\ \tau_{qna} =
\widehat{S}^{na}_q[\nu]_{\nu=0}.
\enn

We define the frequency spectrum as Fourier transformation of
these quantities by relation
\bd
\widehat{M}_n(\nu) = \int_{-\infty}^{+\infty} M_n(\tau) e^{-2\pi
i\nu\tau} d\tau.
\ed
To calculate the Fourier transformation of all quantities we use
package for numerical Fourier transformation from package
"Mathematica". We tabulate functions $M_n(\tau)$ with step
$\triangle = 1/40$ in range $\tau = (0,60)$. The amount of points
$N=2401$. Then the Fourier transform
\bnn
\widehat{M}_n(\nu) &=& \int_{-\infty}^{+\infty} M_n(\tau)
e^{-2i\pi\nu\tau} d\tau = 2 \int_{0}^{+\infty} M_n(\tau) \cos
(2\pi\nu\tau) d\tau\adb\nonumber\\
&=& 2 \Delta \Re \sum_{k=0}^{N-1}M_n(k\Delta)e^{-2\pi i k\nu/N} -
M_n(0)
\enn
is represented as a function of discrete frequencies $\nu =
\fr{k}{N\Delta}\ (k=0,\ldots, N-1)$
\bn\label{DiscreteF}
\widehat{M}_n(\nu) &=& 2 \Delta \Re
\sum_{l=0}^{N-1}M_n(k\Delta)e^{-2\pi i l k/N} - M_n(0).
\en

Let us summarize here some analytical results. In the limit $q\to
1$ the formulas \Ref{QS} transform to the equations \Ref{Shannon}.
For this reason we may refer for entropies \Ref{Shannon} as
entropies \Ref{QS} at the point $q=1$. The entropy $S^n_q$
\Ref{QS} amounts to maximum value, $(2^{1-q}-1)/(1-q)$, at the
point $|M_n(t)|^2 = \fr 12$. For small values of $|M_n(t)|^2$ the
dependence of entropies $S^n_q,\ S^{nc}_q$ changes drastically on
value of $q$. For $q \ll 1$ we have
\be
S_q \approx |M_0(t)|^{2q}, \ S^{c}_q \approx |M_0(t)|^{2q},\
S^{a}_q \approx |M_0(t)|^{2}.\label{SqApp}
\ee
These expression are valid for all cases: $|M_0(t)|^2 \ll
q,|M_0(t)|^2 \gg q$ and $|M_0(t)|^2 \sim q$. For comparison we
have for small values of $|M_0(t)|^2$: $S \approx -|M_0(t)|^{2}
\ln |M_0(t)|^{2}$. Therefore $S^{a}_q \ll S^{c}_q,S_q,S$ and we
observe that decreasing of $q$ does as magnifier for small values
of $|M_0(t)|^2$.

For $q\gg 1$ and still $|M_0(t)|^2\ll 1$ we have to consider three
cases $q|M_0(t)|^2\ll 1,\ q|M_0(t)|^2\sim 1$ and $\ q|M_0(t)|^2\gg
1$. We have correspondingly
\bnn
S_q &\approx& |M_0(t)|^{2},\ S^{c}_q \approx \fr 1q|M_0(t)|^{2},\
S^{a}_q \approx |M_0(t)|^{2},\ \left[q|M_0(t)|^2\ll 1\right] \adb\\
S_q &\approx& -\fr 1q ((1-|M_0(t)|^{2})^q -1),\ S^{c}_q \approx
\fr
1q|M_0(t)|^{2},\ S^{a}_q \approx -\fr 1q ((1-|M_0(t)|^{2})^q -1),\
\left[q|M_0(t)|^2\sim 1\right] \adb\\
S_q &\approx& \fr 1q,\ S^{c}_q \approx \fr 1q|M_0(t)|^{2},\
S^{a}_q \approx \fr 1q,\ \left[q|M_0(t)|^2\gg 1\right]\adb.
\enn
We observe that for sufficiently large $q$ and small $|M_0(t)|^2
\ll 1$ we have: $S_q,S^{c}_q,S^{a}_q \ll S$. Therefore we observe
that increasing of $q$ do as demagnification lens for small values
of $|M_0(t)|$. We note that the entropies equal to zero for
arbitrary $q$ for zero value of $M_0(t)$ as well as for
$M_0(t)=1$. It is easy to see, if the TCF $M_0(t)$ possesses an
extremum at some time $t_0$ the entropies have extremum, too. The
structure of extremums of TCF generates the same structure of
extremums of entropies. After this general frameworks let us
consider specific models.

\section{The motion of Brownian oscillator}\label{Sec:3}

Let us consider the Brownian oscillator which models the particle
with internal oscillatory degree of freedom with frequency
$\omega_0$. The friction coefficient $\beta$ describes the
movement of the particle in a medium. We may use the general
theory of Zwanzug and Mori \cite{Zwa61,Mor65} for this particle
because its position and velocity are random quantities.

The position and velocity of the Brownian oscillator is subject
for equations
\bn
\frac{dx}{dt}&=&v, \\
\frac{dv}{dt}&=&-2cv-\omega_0^2x+\frac{F(t)+f(t)}{m},
\label{Langeven}
\en
where $c=\beta/2m$, $F(t)$ -- external force, $f(t)$ -- the
Langevin random force. The random quantities $x, v, f$ are
described by the following average values:
\bnn
\langle x \rangle&=&0, \ \langle v \rangle=0, \ \langle f
\rangle=0, \\
\langle x^2  \rangle &=& \frac{T}{m\omega_0^2}, \ \langle v^2
\rangle =\frac{T}{m}, \ \langle xv \rangle=0, \\
\langle x f \rangle &=& 0, \ \langle v f \rangle=0, \ \langle f(t)
f(t')\rangle=2T\beta \delta (t-t'),
\enn
where $T$ -- temperature in the units of the energy, $m$ -- the
mass of Brownian particle.

We use the fluctuation of particle's position $x$
\bd
W_0=x(0)
\ed
as the initial dynamic variable. The next orthogonal dynamic
variables are calculated with help of the following recurrent
relations:
\bn
W_1 &=& \left\{i\hat L-\omega_0^{(0)}\right\}W_0,  \nonumber \\
W_n&=&\left\{i\hat
L-\omega_0^{(n-1)}\right\}W_{n-1}+\Omega_{n-1}^2 W_{n-2},~~n>1,
\label{eq 2.2.45}
\en
where the frequencies $\omega_0^{(n)}$ and $\Omega_n$ are defined
by equations
\be
 \omega_0^{(n)}=\frac{\langle W^*_n
i\hat L W_n\rangle}{\langle |W_n|^2\rangle},~~
\Omega_n^2=\frac{\langle |W_n|^2\rangle}{\langle
|W_{n-1}|^2\rangle}. \label{eq 2.2.46}
\ee
The straightforward calculations give
\bnn
\omega_0^{(0)} &=& \frac{\langle x(0) v(0)\rangle}{\langle
|x(0)|^2 \rangle} = 0, \adb \\
W_1 &=& (i\hat L - \omega_0^{(0)})W_0=v(0), \adb \\
\omega_0^{(1)} &=& \frac{\langle v(0)
\frac{dv(0)}{dt}\rangle}{\langle |v(0)|^2 \rangle}= -2c, \adb \\
\Omega^2_1 &=& \frac{\langle |v(0)|^2\rangle}{\langle |x(0)|^2
\rangle} = \omega_0^2, \adb \\
W_2 &=& \frac{f(0)}{m}.
\enn
Taking into account the equations of motion \Ref{Langeven}, one
obtains the time correlation function (TCF)
\be
M_0(t)=\frac{\langle x(t) x(0)\rangle}{\langle |x(0)|^2
\rangle}=\frac{r_-e^{r_+t}-r_+e^{r_-t}}{r_--r_+}, \label{a(t)}
\ee
where $r_\pm = -\left[c \pm \sqrt {c^2 - \omega_0^2}\right]$.

To find the TCF for next levels we exploit the Zwanzig-Mori
infinite chain of equations given by Eq. \Ref{MainSystem}. The
Laplace-image of initial TCF has the following form:
\bd
\widetilde{M}_0(s)= \frac{s-(r_-+r_+)}{(s-r_-)(s-r_+)}.
\ed
With $n=0$ we obtain from chain \Ref{Laplac} the Laplace-image of
first memory function
\bd
\widetilde{M}_1(s)=\frac{1}{s-(r_-+r_+)}.
\ed
Taking the inverse Laplace transform we obtain the first memory
function
\bd
M_1(t)=e^{(r_-+r_+)t}=e^{-2ct}.
\ed
The calculation in closed form of the memory functions on the next
relaxation levels requires the following detailed elaboration of
the structure of Langevin force $f$. For example, in order to
calculate the memory function $M_2(t)$ we need for frequency
$\Omega_2$ which may be found by setting the mean value of square
of Langevin force. For this reason we restrict our consideration
by first two levels of relaxation.

We consider the TCF of Brownian oscillator in the case of small
damping $p=c/\omega_0 \ll 1$. In this case the TCF $M_0(t)$ has
the following form:
\be
M_0(t) = \fr{\langle x(t)x(0)\rangle}{\langle x(0)x(0)\rangle} =
\cos (2\pi\nu' t) e^{-c|t|}\label{Oscillator}
\ee
and describes the  motion of Brownian oscillator with frequency
$\omega = 2\pi\nu' \gg c = \beta/2m$, where $m$ is the mass of
oscillator, and $\beta$ is the friction coefficient of Brownian
particle \cite{ResDeL77}.

There are two parameters $\nu'$ and $c$ which characterize the
frequency oscillation and relaxation damping of TCF, respectively.
For simplicity we consider dimensionless time $t$, frequency of
oscillation $\nu'$ and damping parameter $c$. The Fourier
transform of square of this TCF has the following form
\bd
\widehat{M_0^2}(\nu) = \fr c{2(c^2 + \pi^2 \nu^2)} + \fr c{4(c^2 +
\pi^2 (\nu - 2\nu')^2)} + \fr c{4(c^2 + \pi^2 (\nu +2\nu')^2)}.
\ed
We consider frequency spectrum of square of TCF because the
entropies are expressed in terms of square of TCF. Therefore there
are three maximums at $\nu = 0,\nu = \pm 2\nu'$. In the limit of
zero damping $c\to 0$ we obtain
\bd
\widehat{M_0^2}(\nu) = \fr 12 \delta(\nu) + \fr 14 \delta(\nu -
2\nu') + \fr 14 \delta(\nu +2\nu')
\ed
by using well-known formula:
\bd
\pi\delta(x) = \lim_{\varepsilon \to 0} \fr \varepsilon{x^2 +
\varepsilon^2}.
\ed
In this case the spectrum consists of three lines at $\nu = 0,\nu
= \pm 2\nu'$.

\begin{figure}[ht]
\hspace*{-1cm}\centerline{\epsfxsize=6truecm\epsfbox{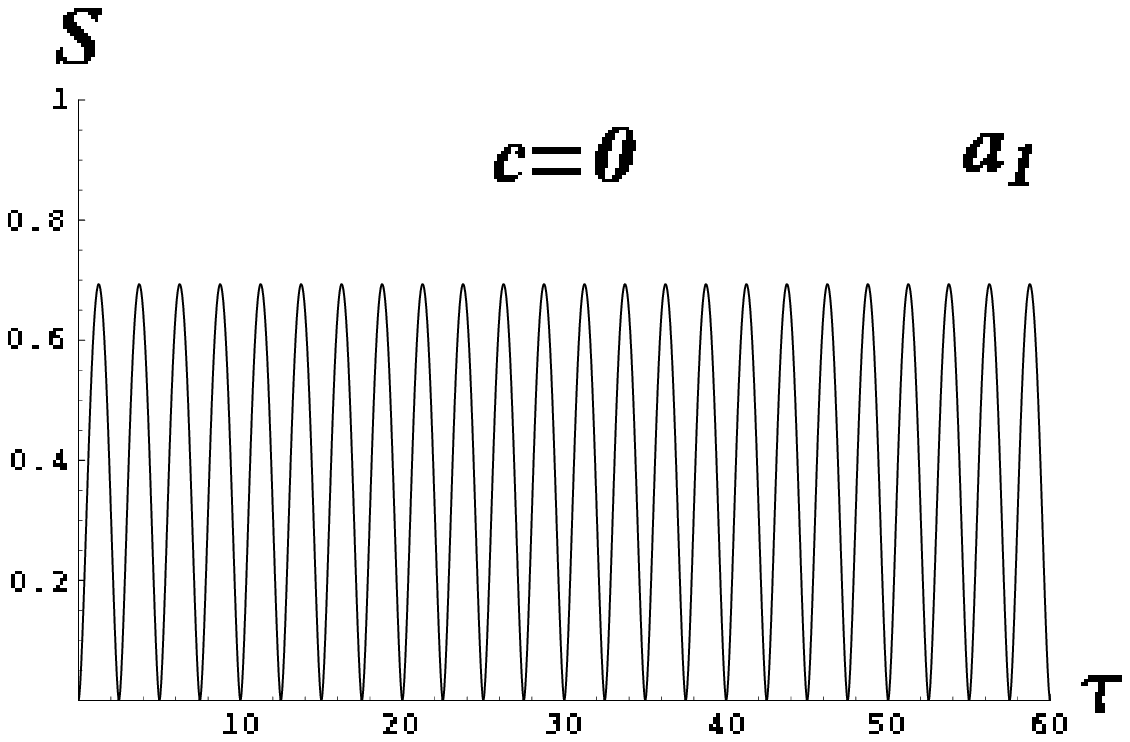}%
\epsfxsize=6truecm\epsfbox{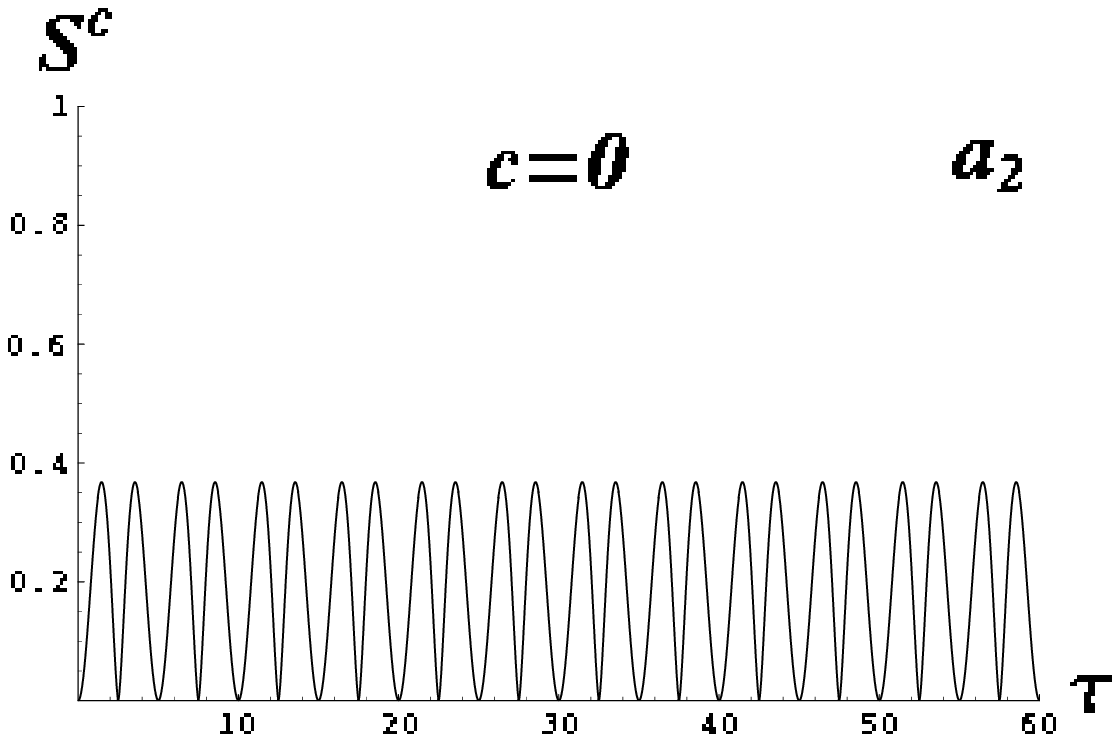}%
\epsfxsize=6truecm\epsfbox{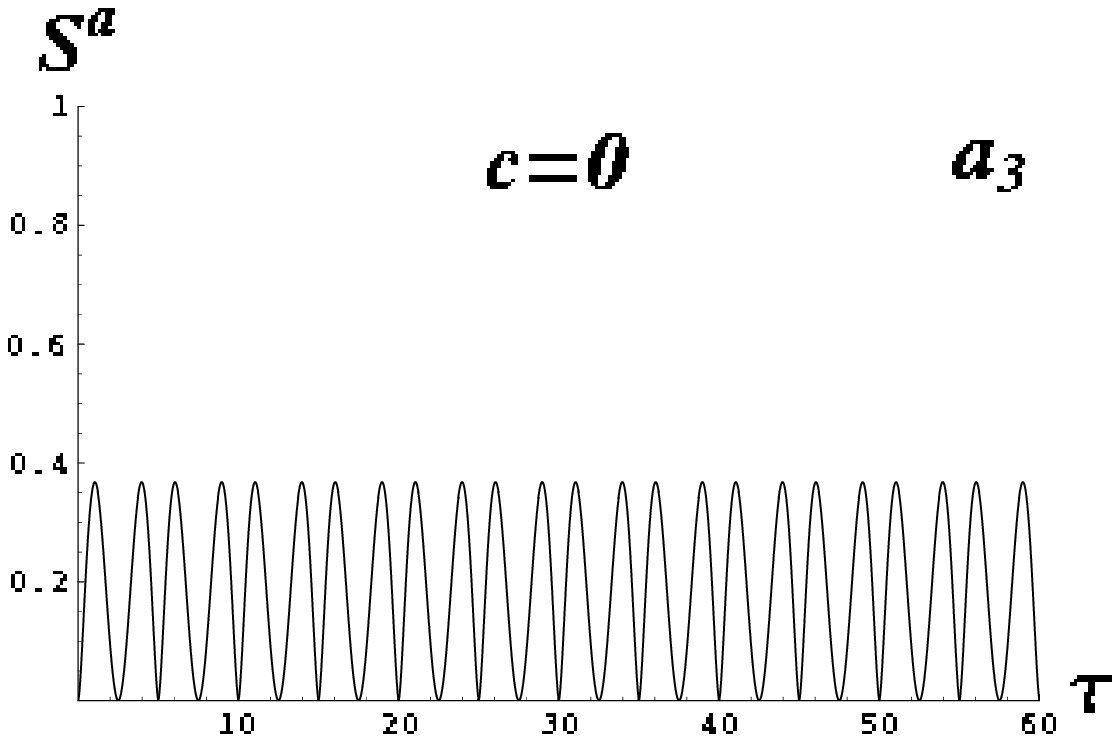}}
\hspace*{-1cm}\centerline{\epsfxsize=6truecm\epsfbox{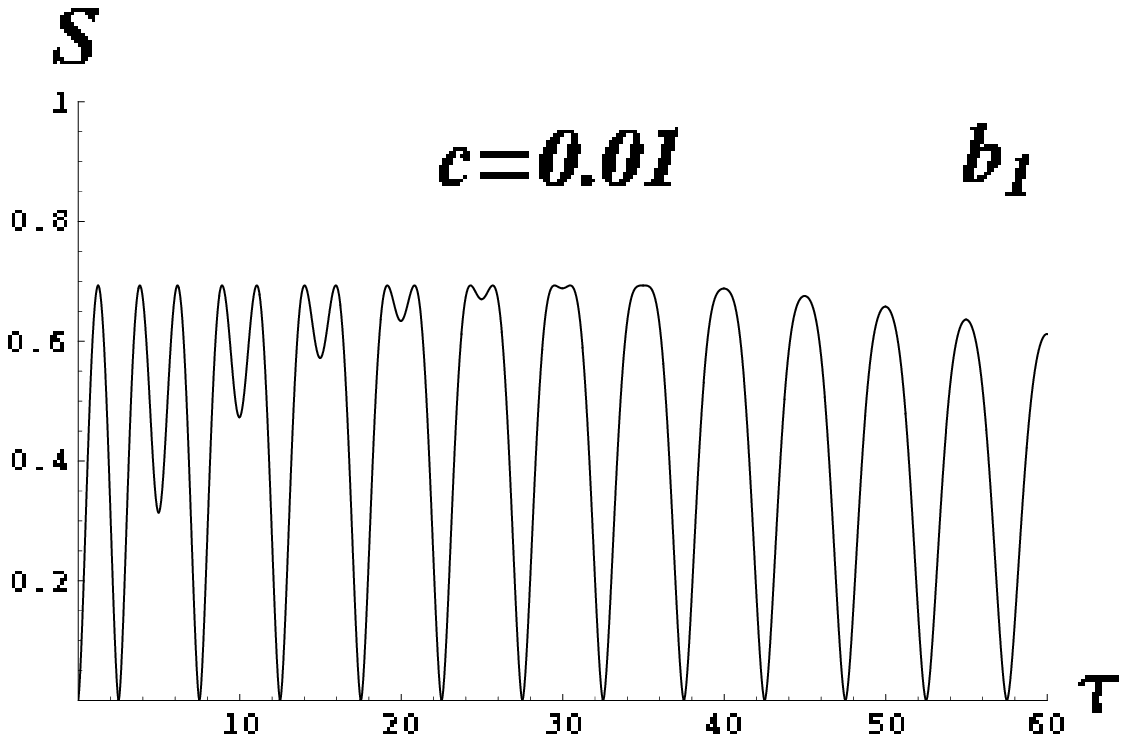}%
\epsfxsize=6truecm\epsfbox{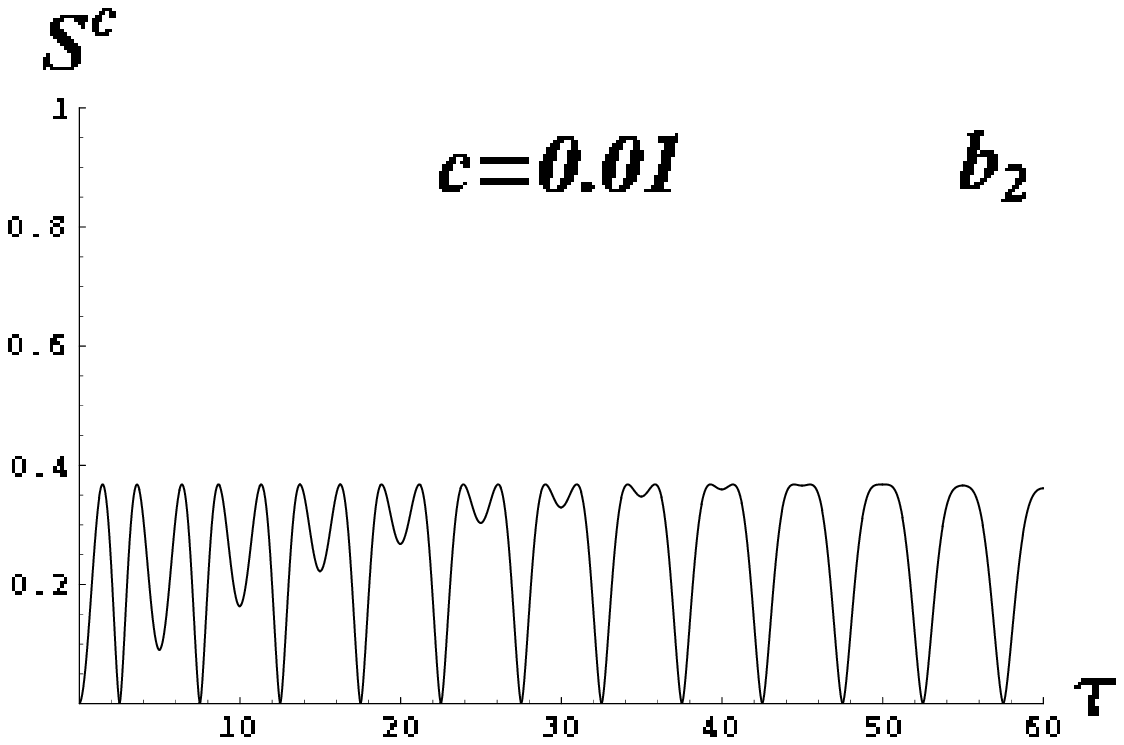}%
\epsfxsize=6truecm\epsfbox{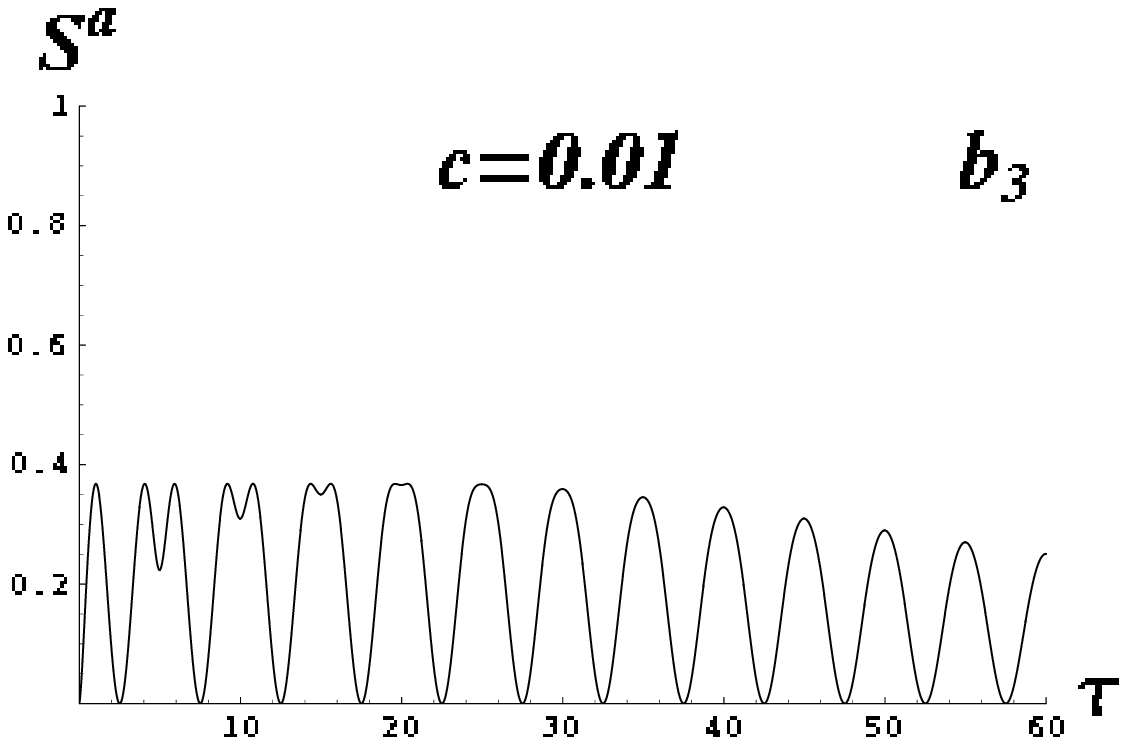}}
\hspace*{-1cm}\centerline{\epsfxsize=6truecm\epsfbox{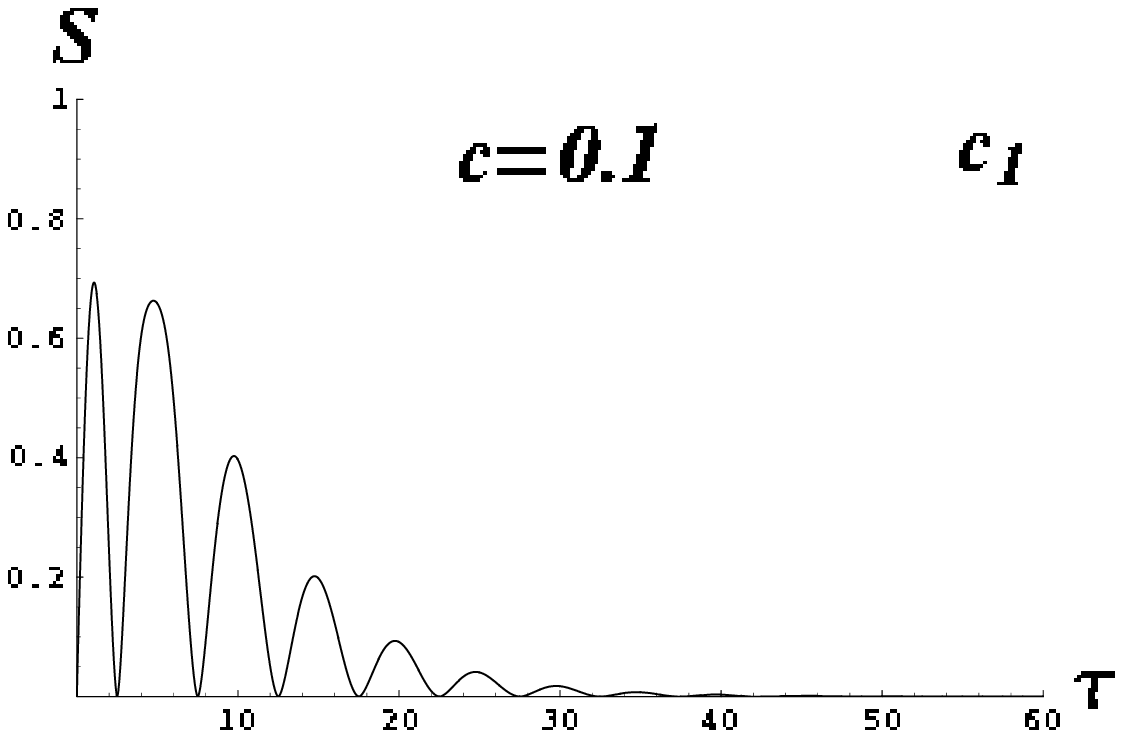}%
\epsfxsize=6truecm\epsfbox{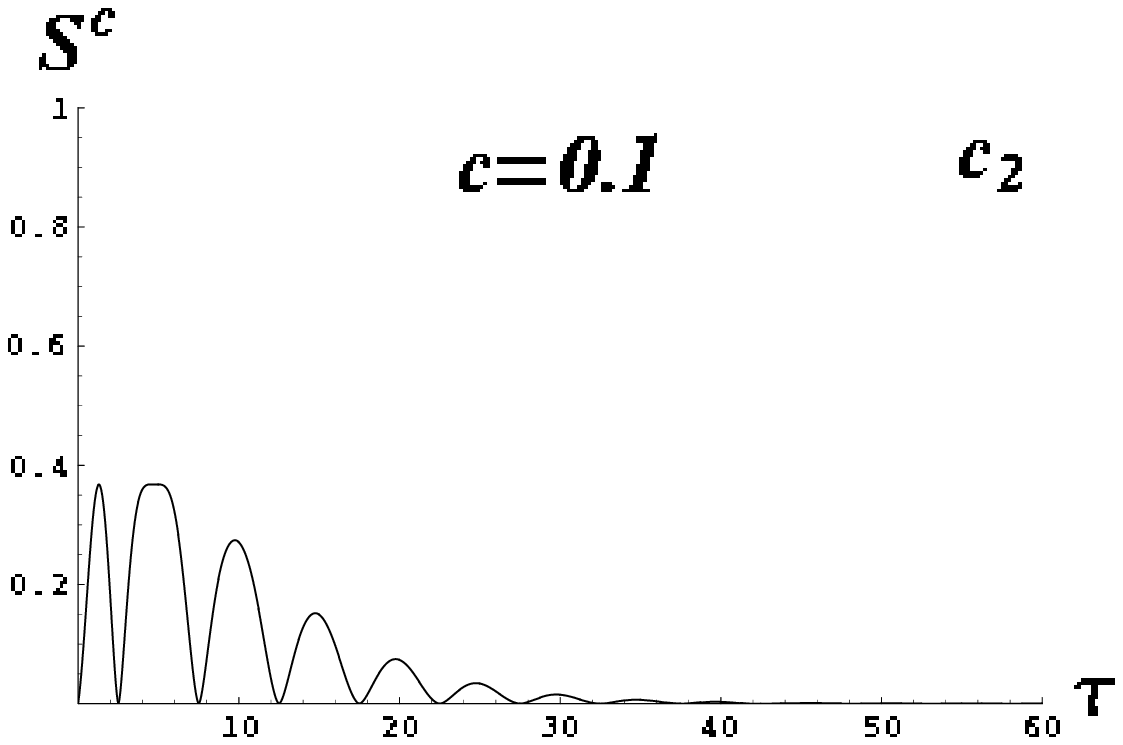}%
\epsfxsize=6truecm\epsfbox{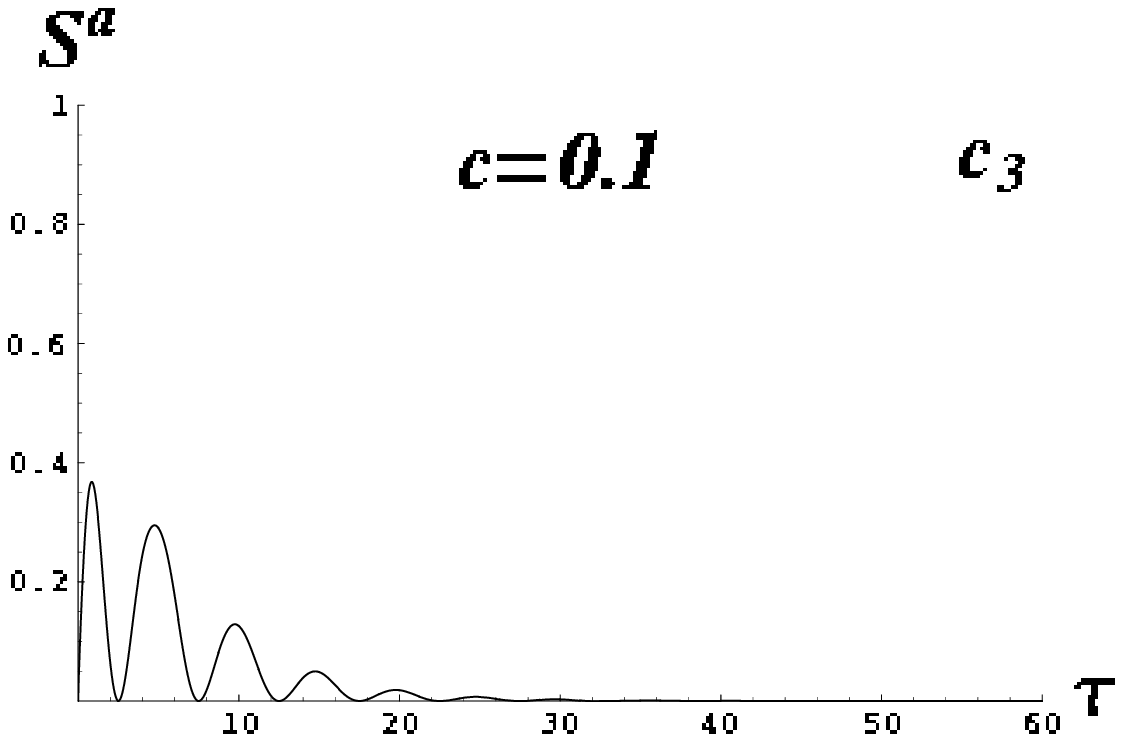}}
\hspace*{-1cm}\centerline{\epsfxsize=6truecm\epsfbox{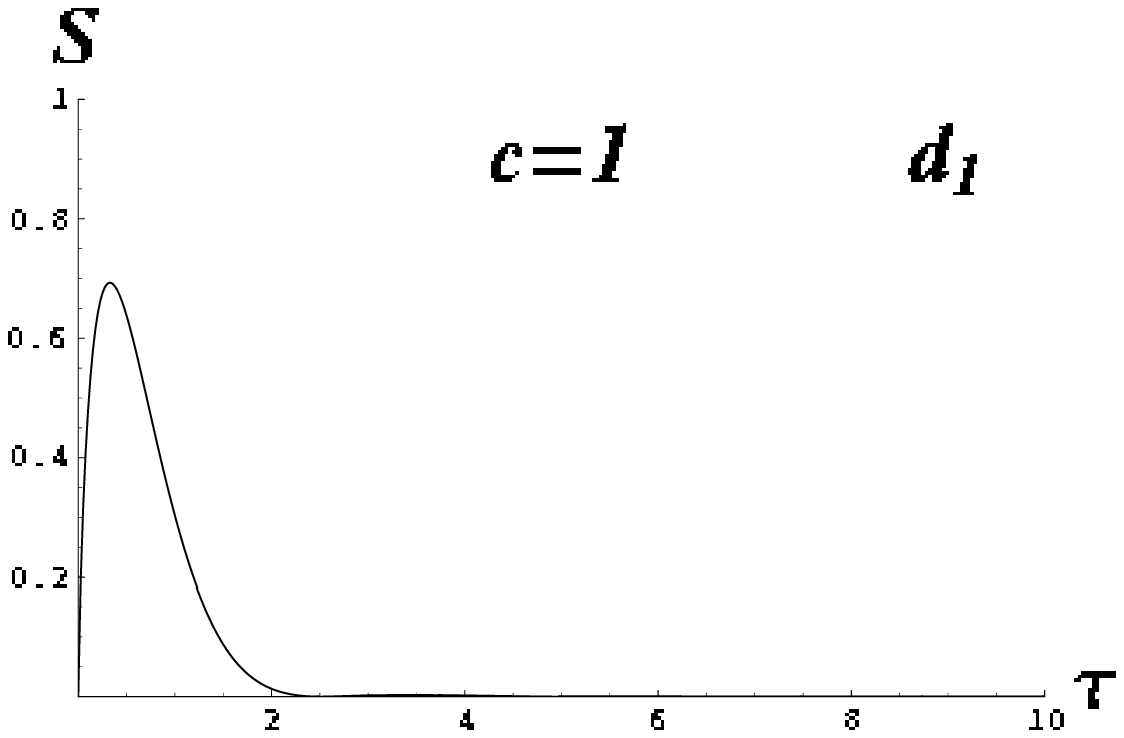}%
\epsfxsize=6truecm\epsfbox{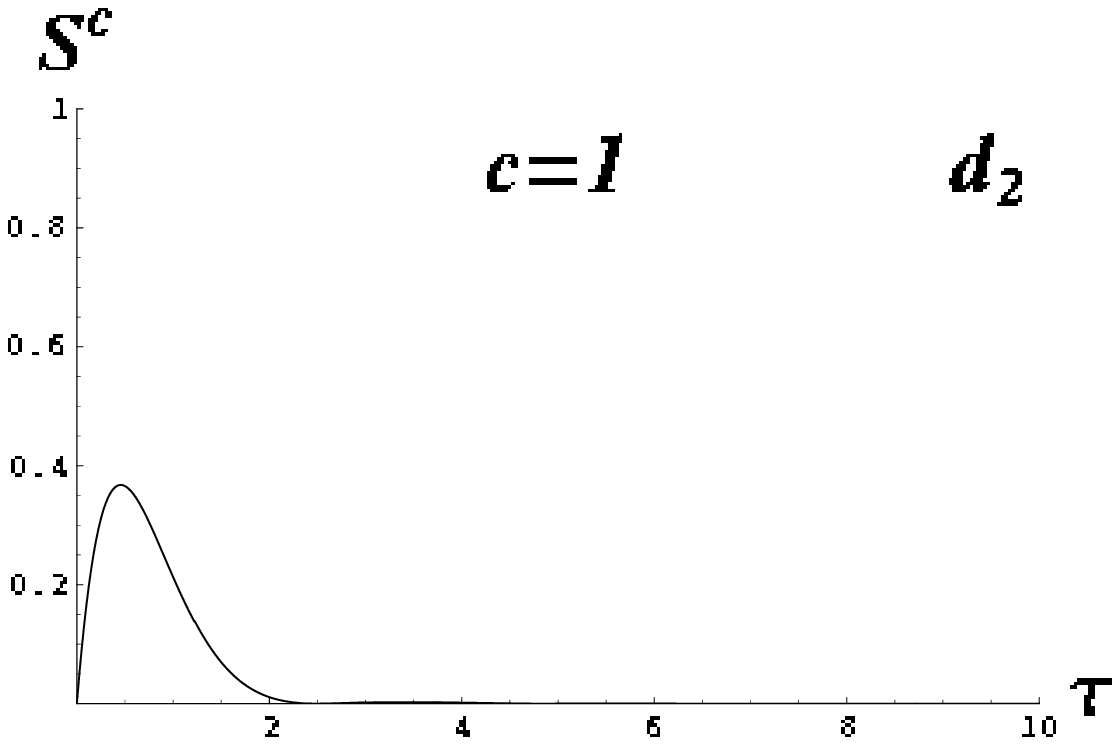}%
\epsfxsize=6truecm\epsfbox{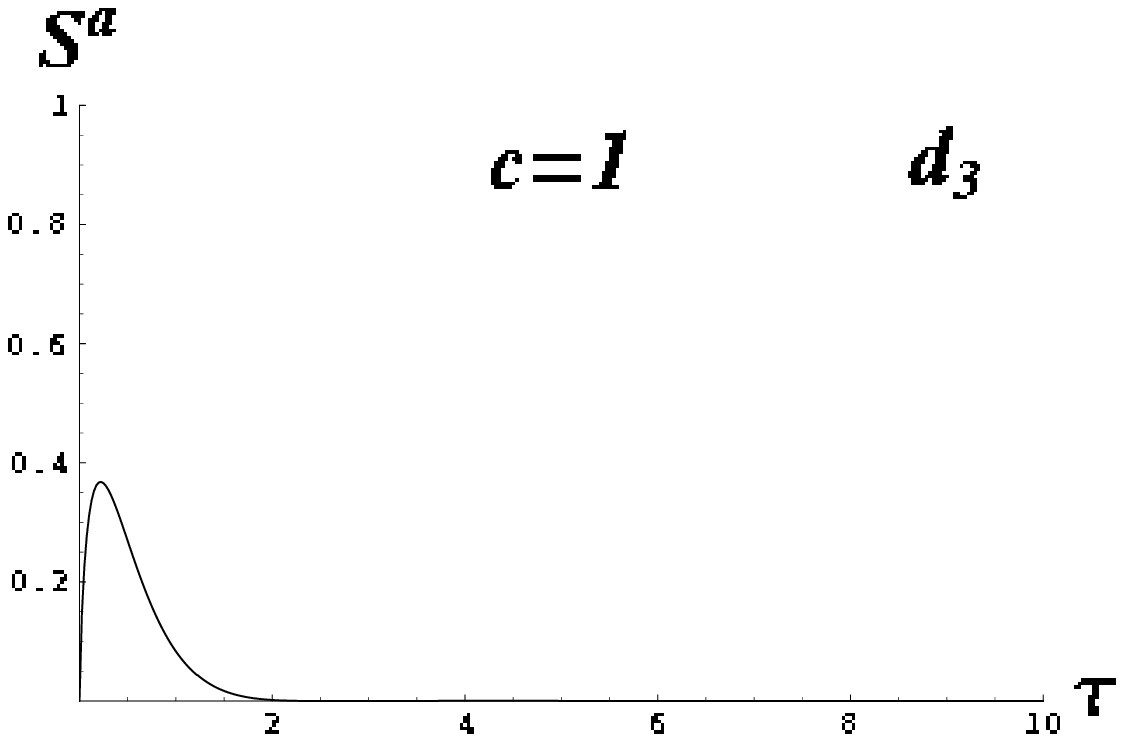}} \caption{The plots of
entropies \Ref{Shannon} for frequency $\nu' = 0.1$ and
$c=0(a),0.01(b),0.1(c),1(d)$ ($\nu'$ and $c$ are oscillation and
damping parameters of TCF \Ref{Oscillator}). It is seen that the
time behavior of total $S$ and single channels $S^c$ and $S^a$ of
DTEs for various damping regimes reveal a stochastic ordering of
time correlation.}\label{Toy1}
\end{figure}

\begin{figure}[ht]
\hspace{-1cm}\centerline{\epsfxsize=6truecm\epsfbox{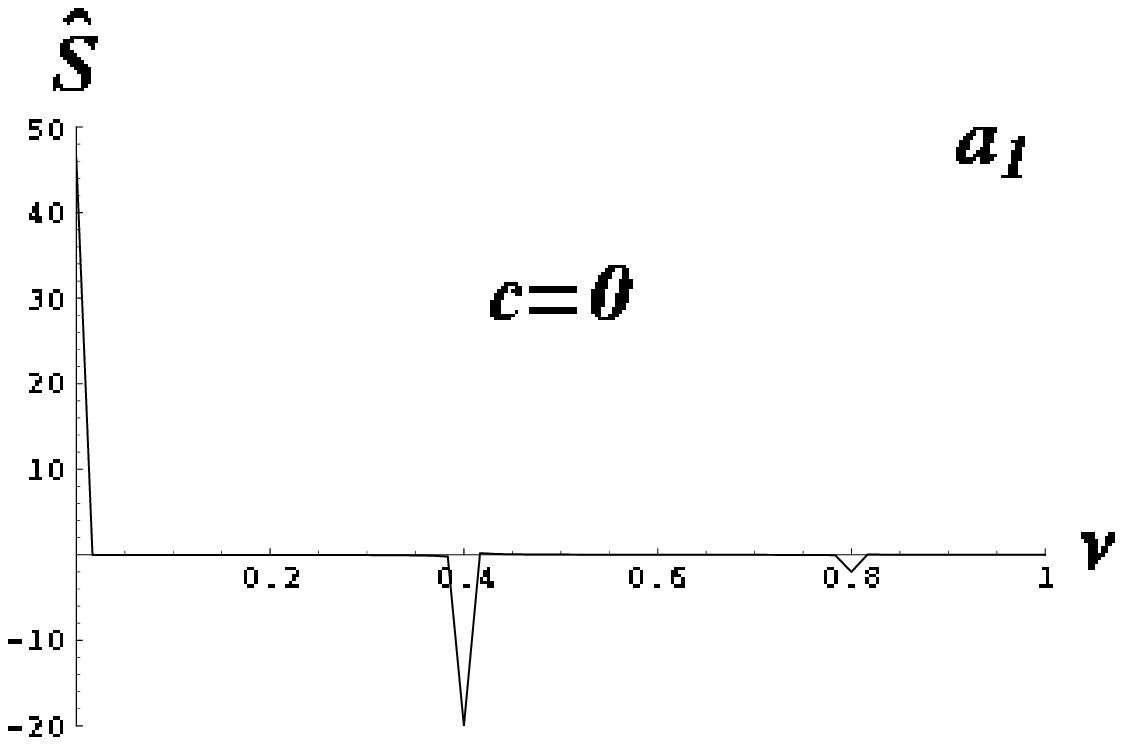}%
\epsfxsize=6truecm\epsfbox{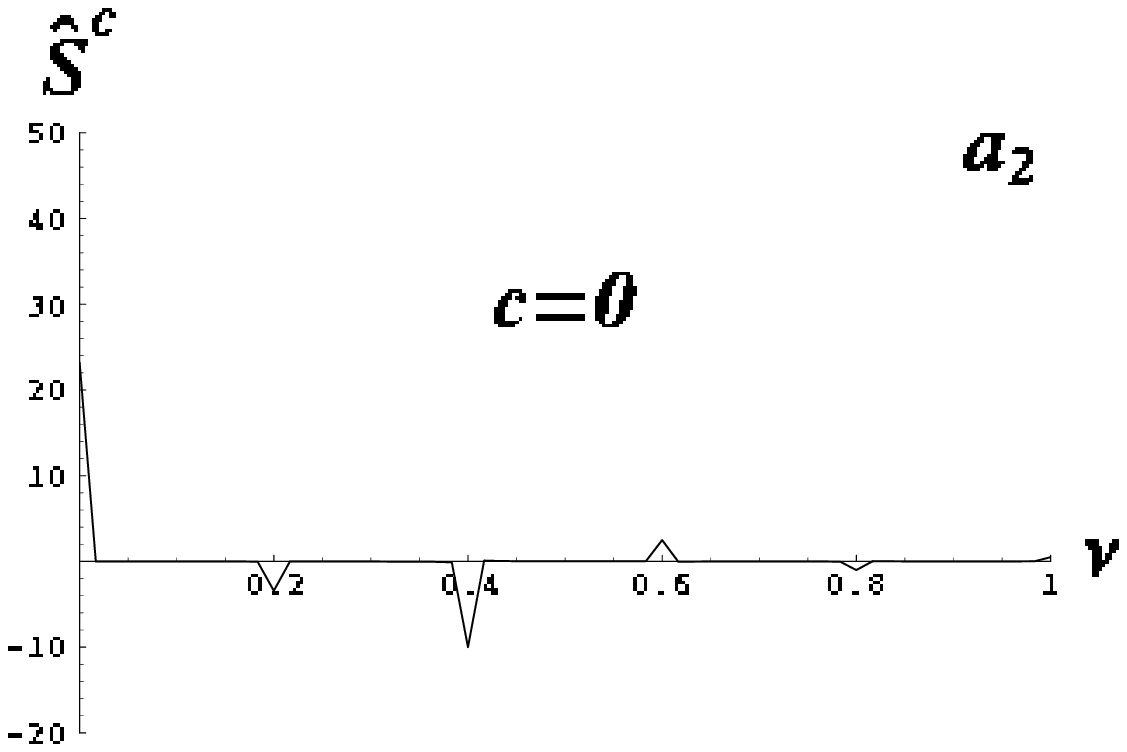}%
\epsfxsize=6truecm\epsfbox{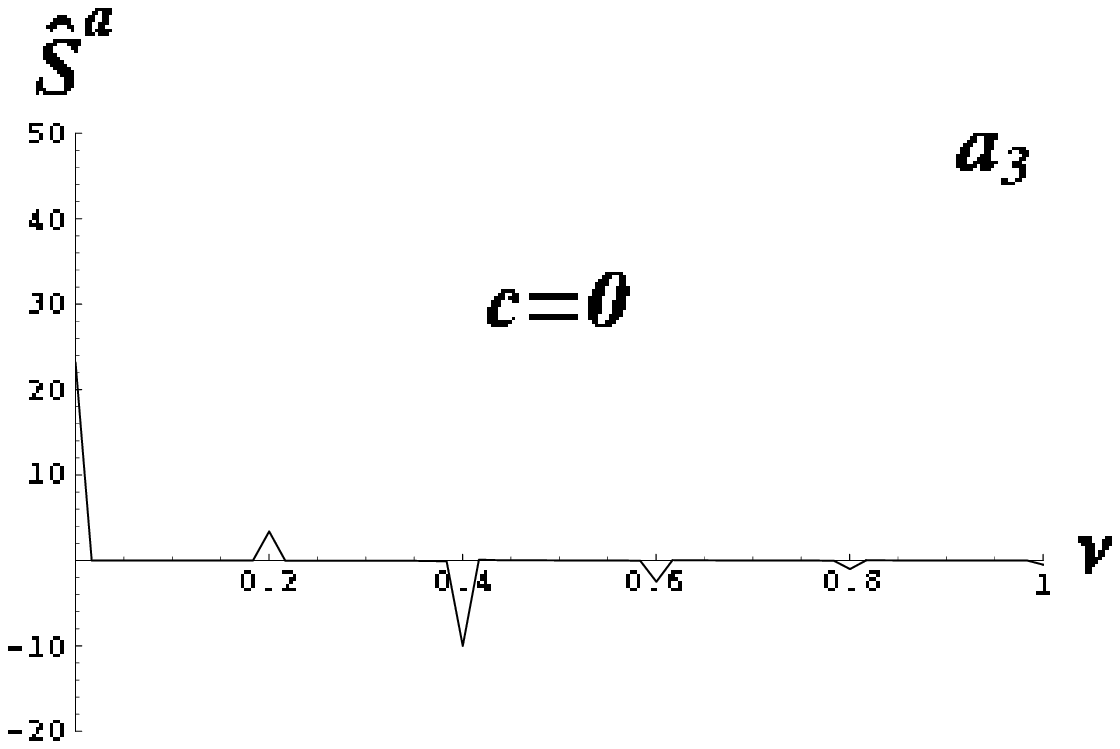}}
\hspace*{-1cm}\centerline{\epsfxsize=6truecm\epsfbox{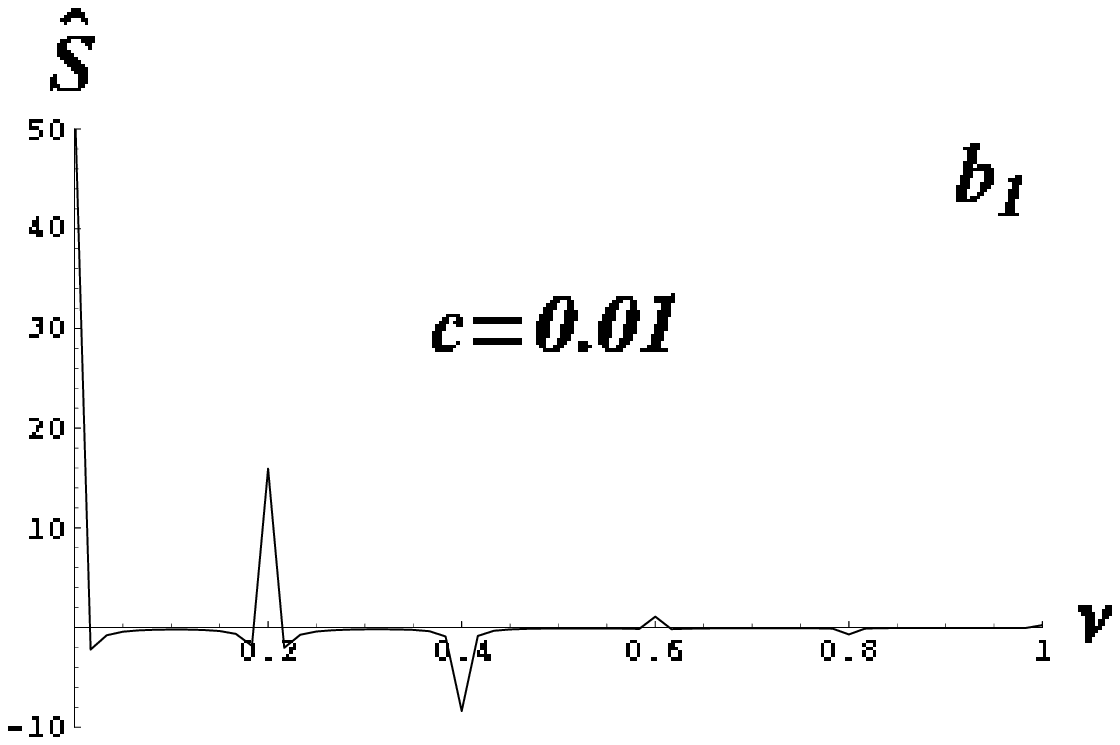}%
\epsfxsize=6truecm\epsfbox{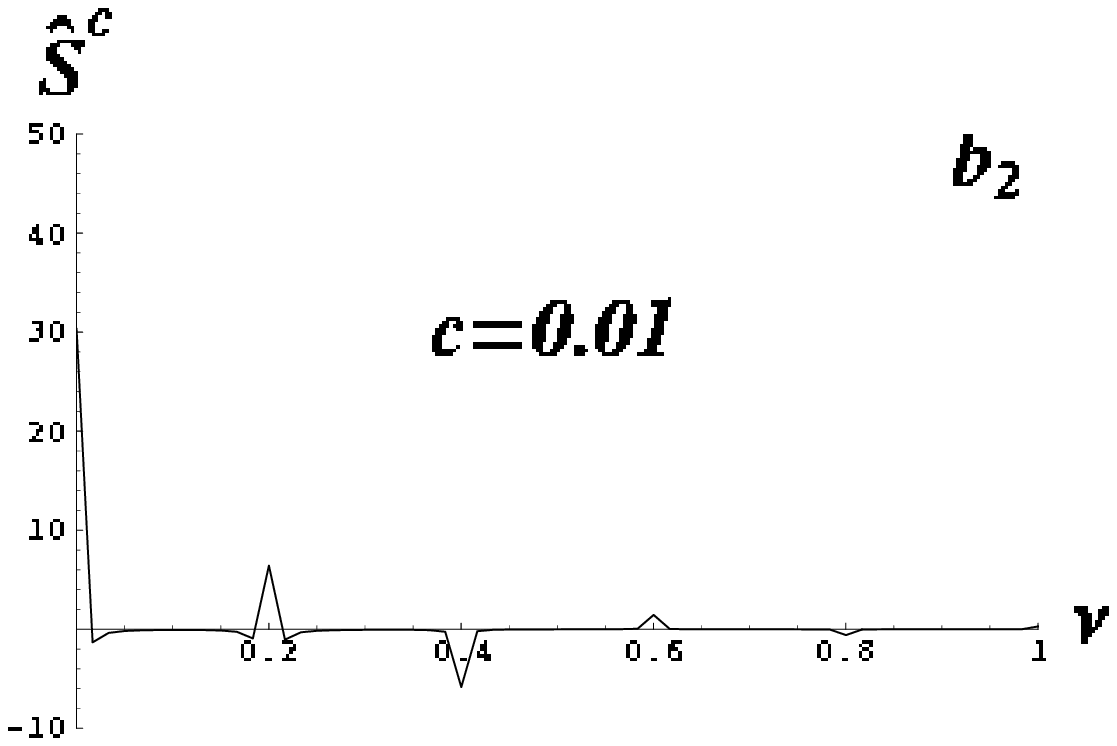}%
\epsfxsize=6truecm\epsfbox{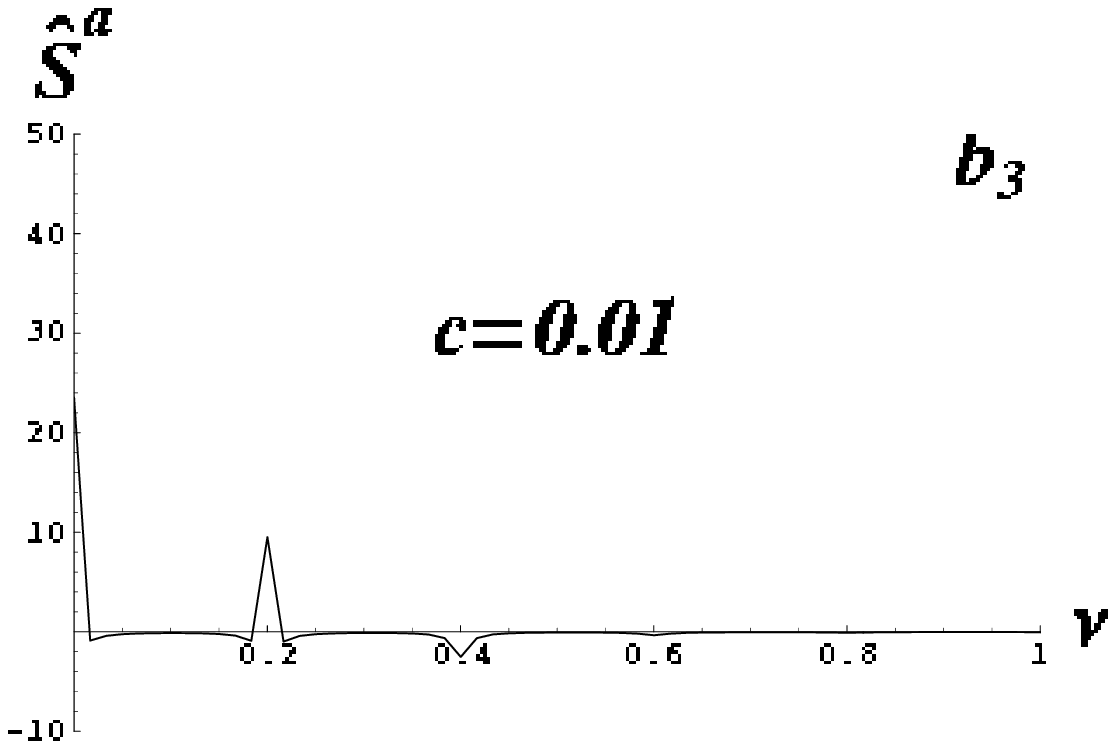}}
\hspace*{-1cm}\centerline{\epsfxsize=6truecm\epsfbox{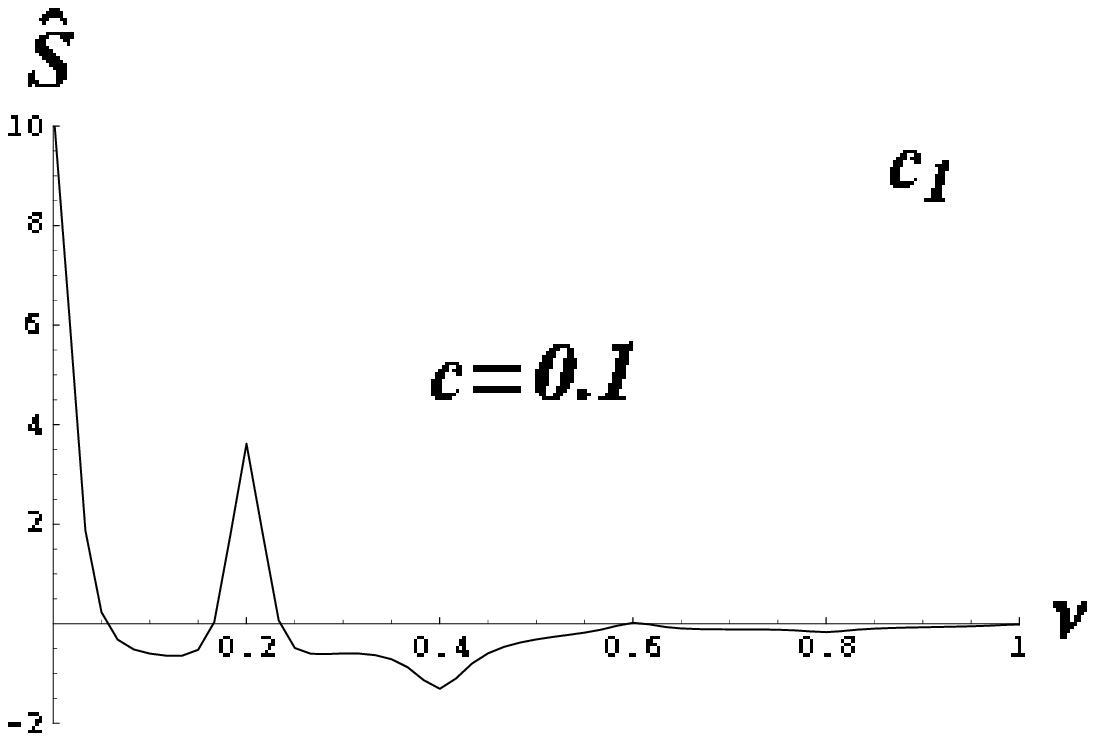}%
\epsfxsize=6truecm\epsfbox{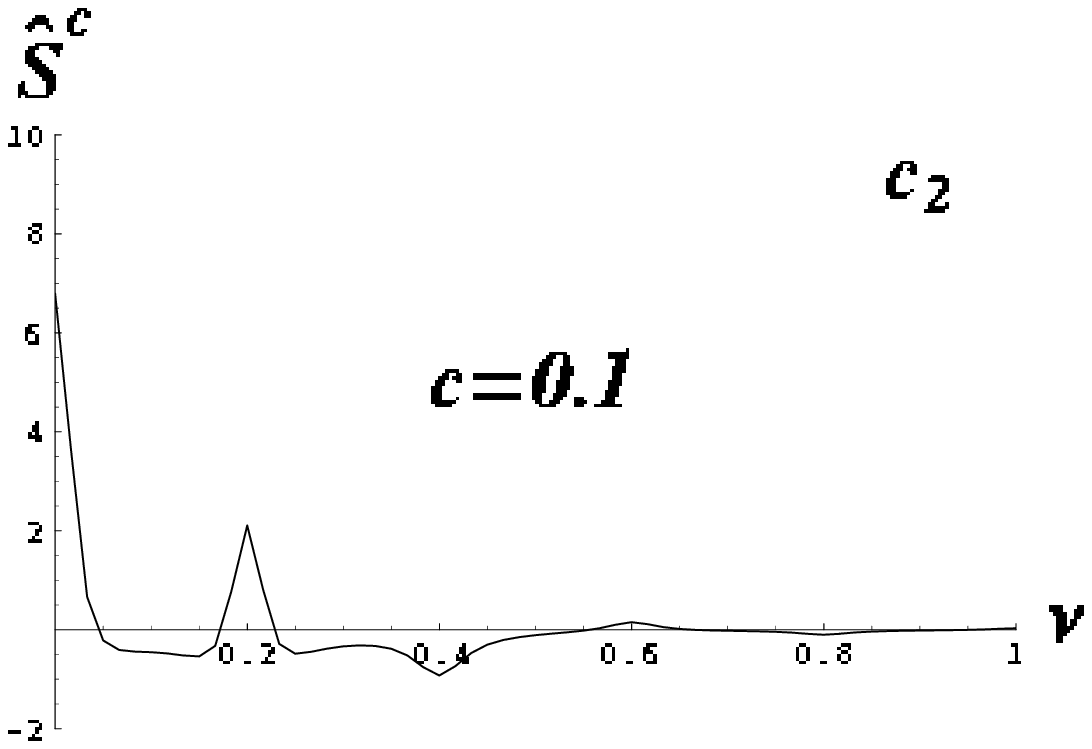}%
\epsfxsize=6truecm\epsfbox{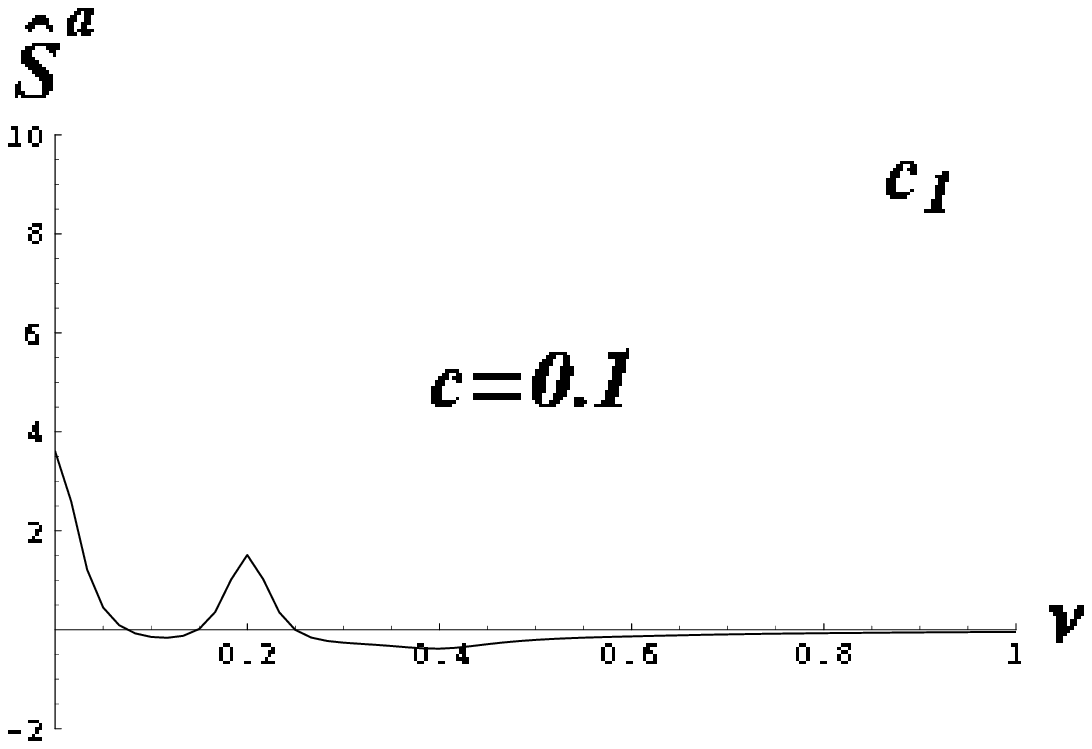}}
\hspace*{-1cm}\centerline{\epsfxsize=6truecm\epsfbox{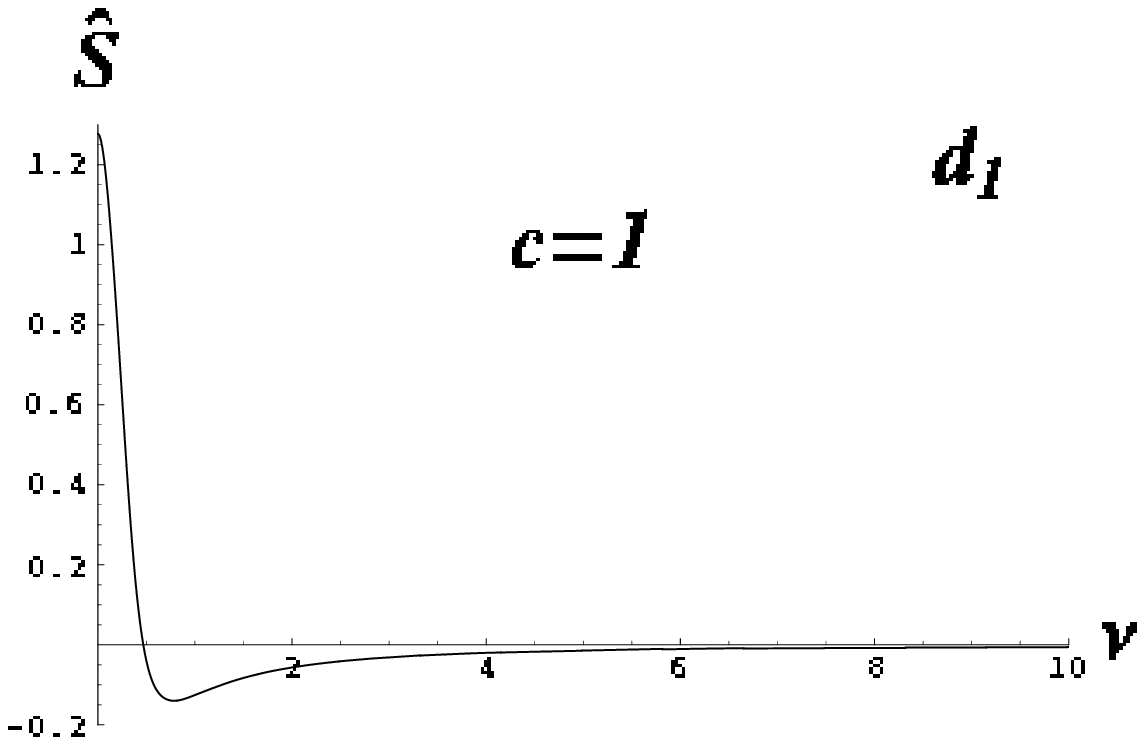}%
\epsfxsize=6truecm\epsfbox{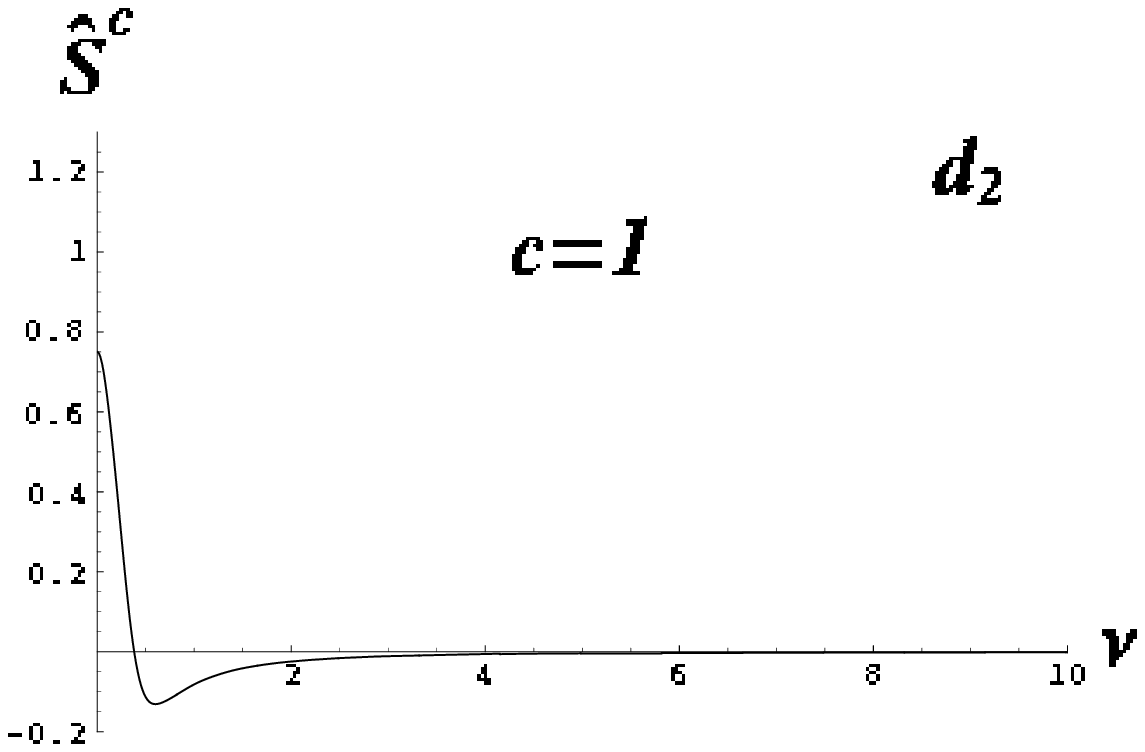}%
\epsfxsize=6truecm\epsfbox{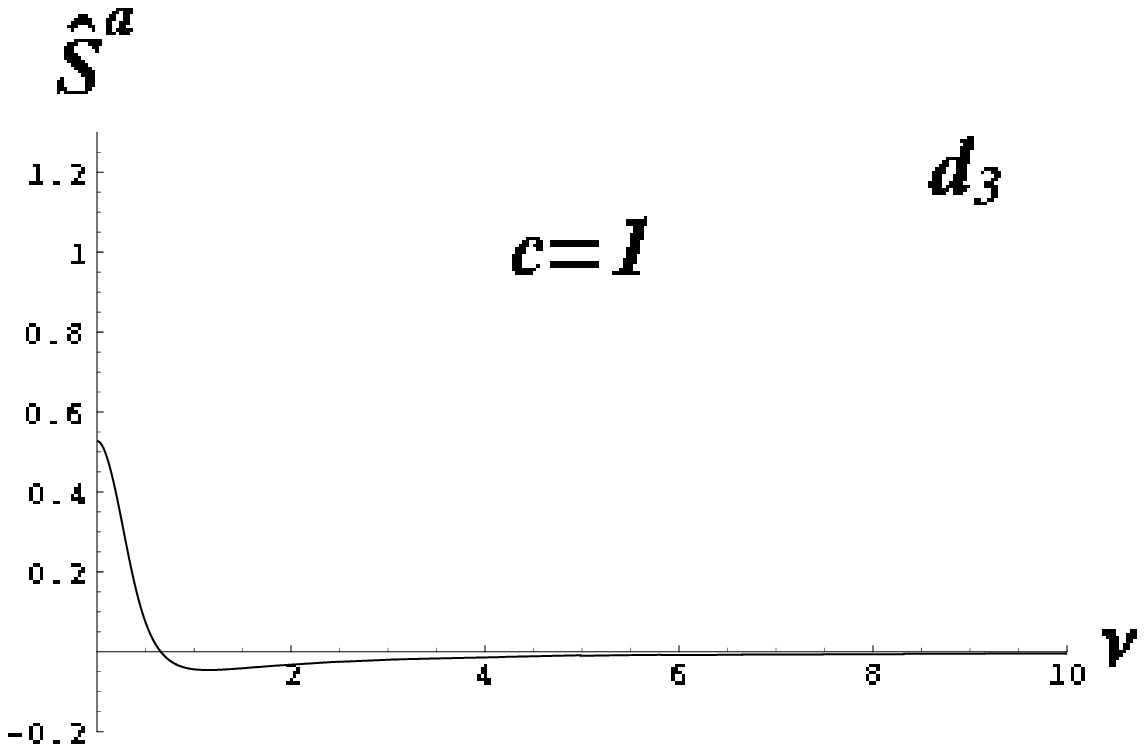}} \caption{The plots of
spectra of entropies \Ref{Shannon} for $\nu' = 0.1$ and
$c=0(a),0.01(b),0.1(c),1(d)$. One can see that a stochastic
ordering of time correlation in previous Fig.\ref{Toy1} reduces to
an appearance of specific peculiarities in low frequency region.
Weak damping ($c=0.1$) results in amplification of characteristic
frequency peaks at $\nu = 0.2$ and $\nu = 0.4$ whereas zero
damping ($c=0$) leads to a dissapearance of specific frequency
peaks.}\label{Toy2}
\end{figure}

In Fig.\ref{Toy1} we reproduce the entropies \Ref{Shannon} and in
Fig.\ref{Toy2} we show the spectra of them for $\nu' = 0.1$ and
for $c=0,0.01,0.1,1$. There is a peak in spectra at double
frequency $\nu = 2\nu' = 0.2$ for arbitrary small but non-zero
damping $c$. For zero damping $c=0$ this peak disappears in total
entropy. There are peaks in $\widehat{S}^c$ and $\widehat{S}^a$ at
$\nu = 0.2$ but with opposite sign.  We note that the increasing
of the damping leads to smearing the fine structure of entropies.

\begin{figure}
{\hspace{-1.5cm}\centerline{\epsfxsize=6.5truecm\epsfbox{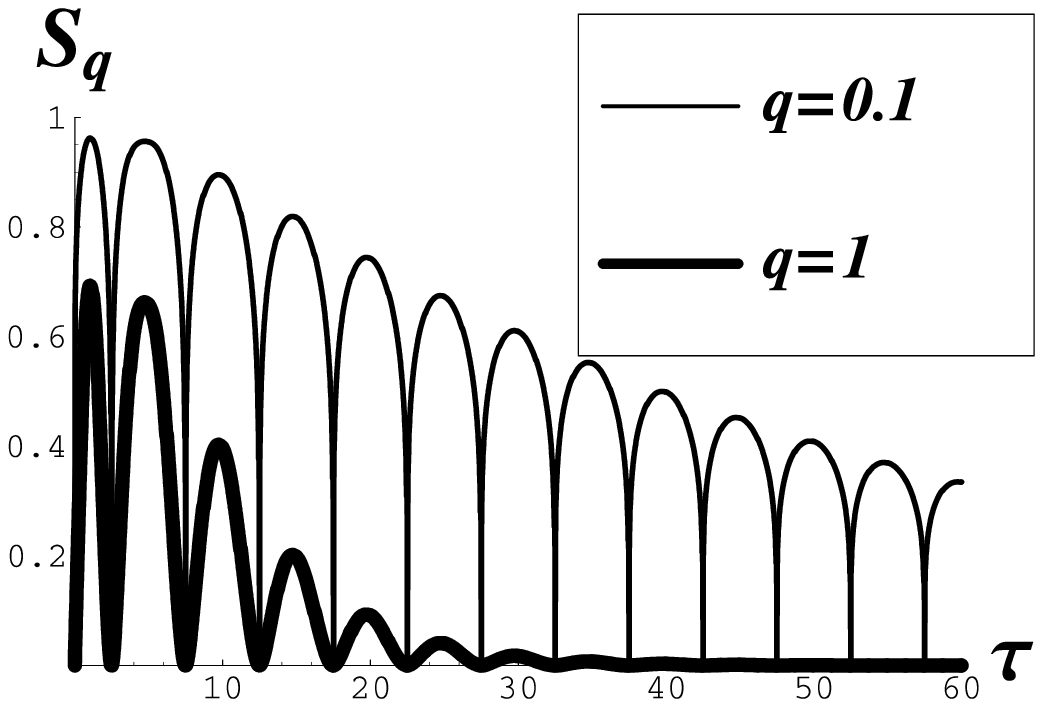}%
\epsfxsize=6.5truecm\epsfbox{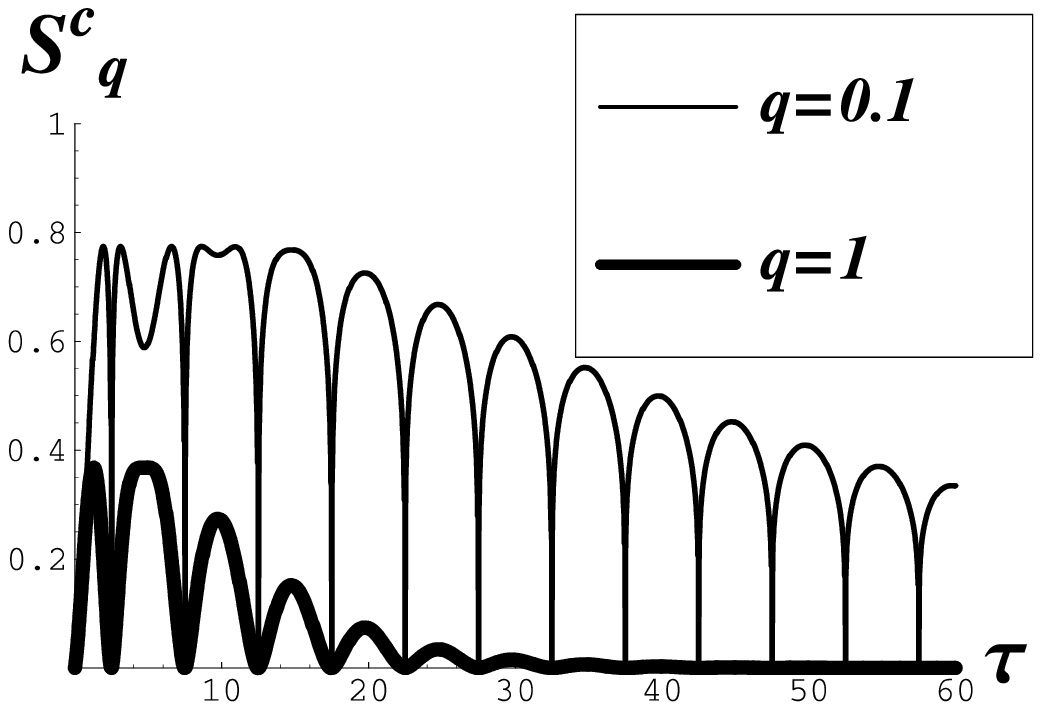}%
\epsfxsize=6.5truecm\epsfbox{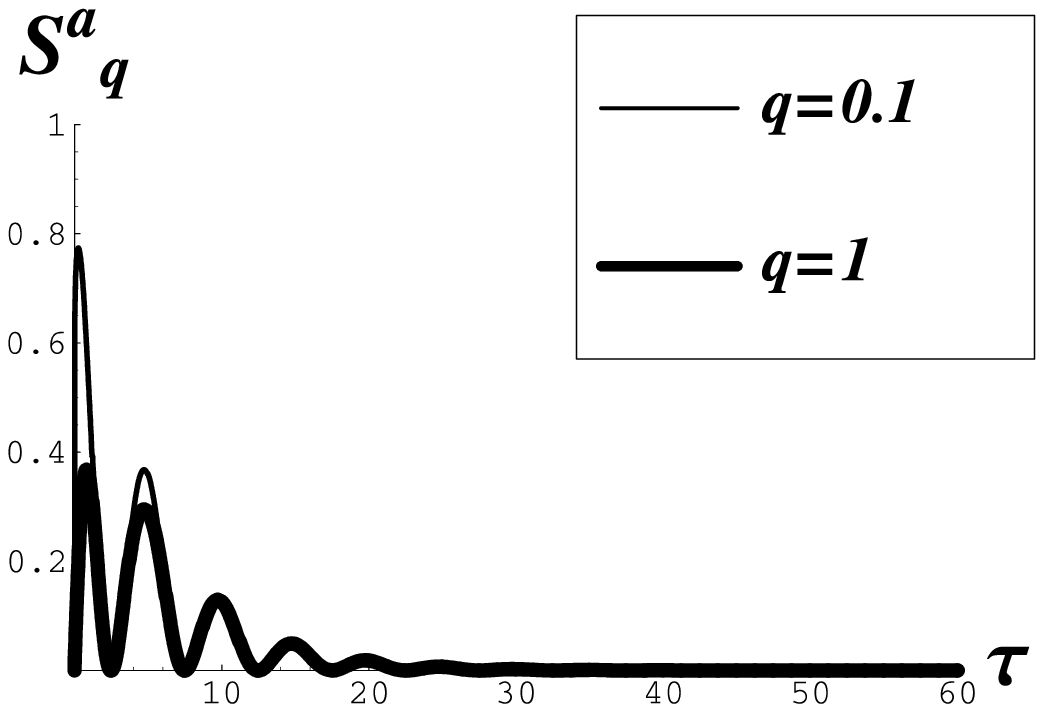}}}

\vspace{-8cm}\hspace{-1.5cm}\centerline{\epsfxsize=6.5truecm\epsfbox{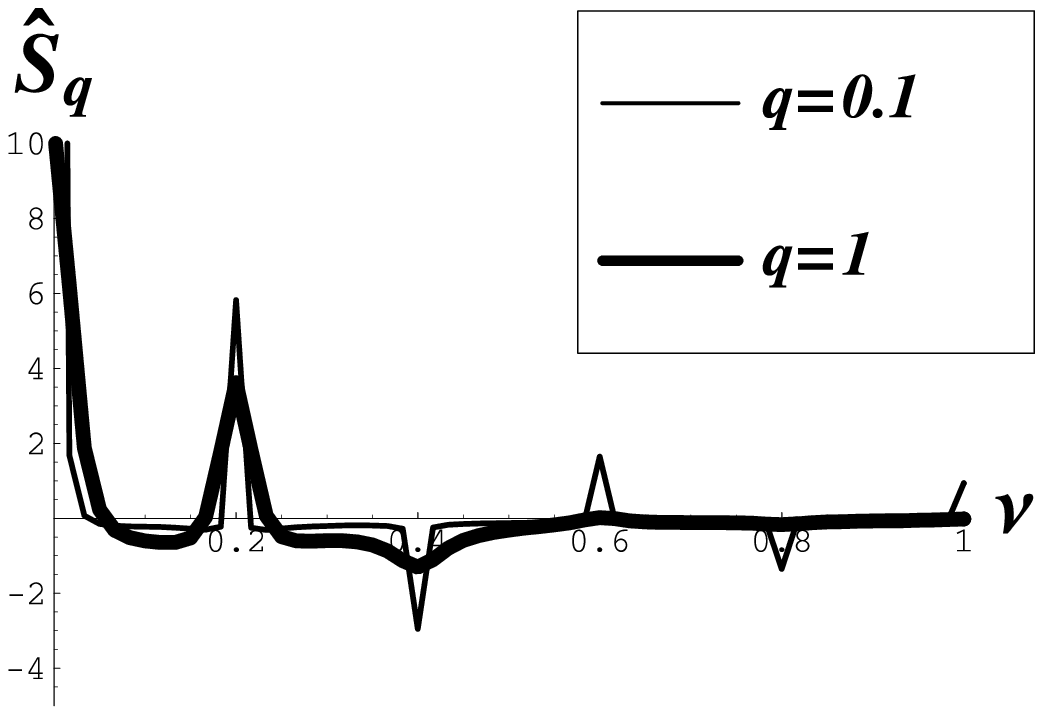}%
\epsfxsize=6.5truecm\epsfbox{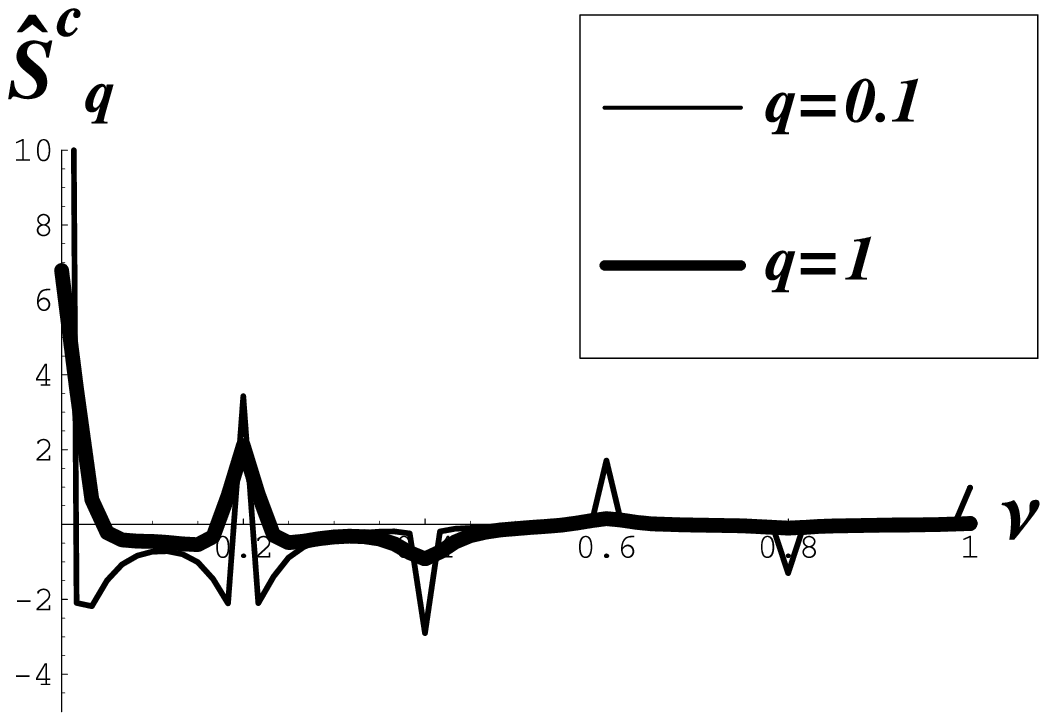}%
\epsfxsize=6.5truecm\epsfbox{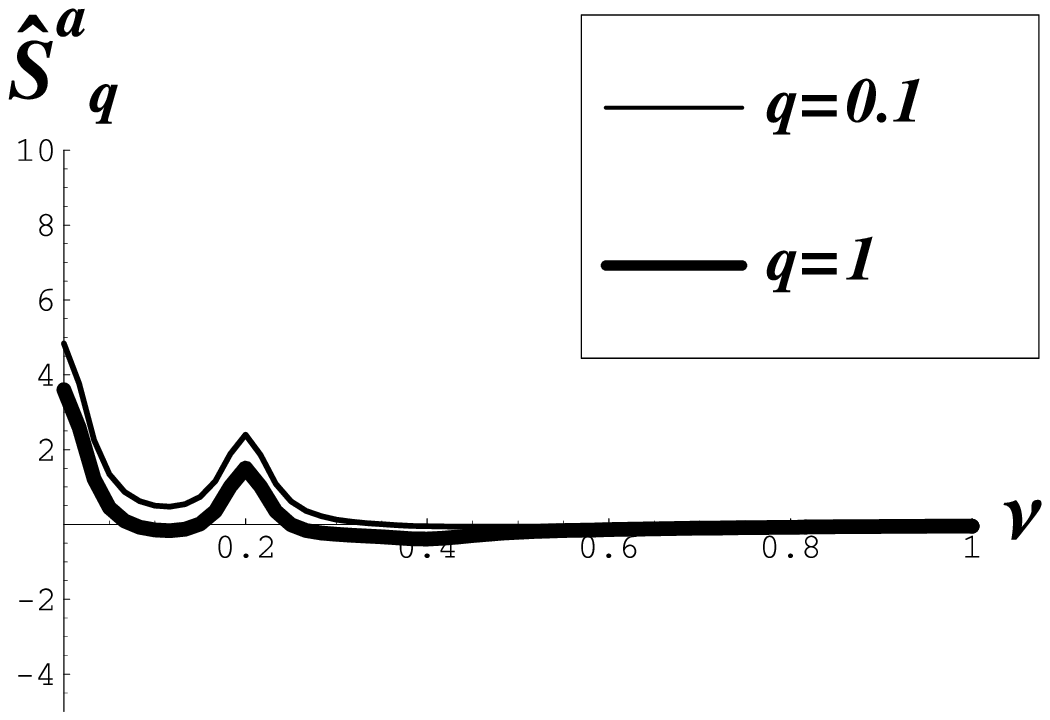}}

\vspace{-8cm} \caption{The plots of entropies \Ref{QS} for $q=0.1$
and \Ref{Shannon} ($q=1$) and their spectra for $\nu' = 0.1,
c=0.1$. Comparison of dynamic Shannon ($q=1$) and Tsallis
($q=0.1$) entropies shows the amplification of DTE for small
values of TCF. Therefore one can conclude that DTE acts as a
magnifier for small values of TCF.} \label{Toy3}
\end{figure}

Calculation the entropies \Ref{QS} shows that the small values of
$q$ "works"\/ as non-linear magnifier for small value of $M_0^2$.
To illustrate this fact we reproduce in Fig. \ref{Toy3} the
entropies \Ref{QS} and \Ref{Shannon} for $q=1,0.1,\ \nu' =0.1,\
c=0.1$. First of all, as expected from Eq. \Ref{SqApp} the smaller
entropy the greater magnification. Second, the entropy $S^a_q$
does not change sufficiently. The great variation prove $S_q$ and
$S^c_q$. The decreasing of $q$ leads to increasing the value of
peaks for great frequencies and makes peaks more sharp. The small
$q$ makes better sharpness of frequency spectra. This is not the
case for $S^a_q$. It does not change great.

\section{The motion of Brownian oscillator with noise}\label{Sec:4}

\begin{figure}
\hspace{-1.5cm}\centerline{\epsfxsize=6.5truecm\epsfbox{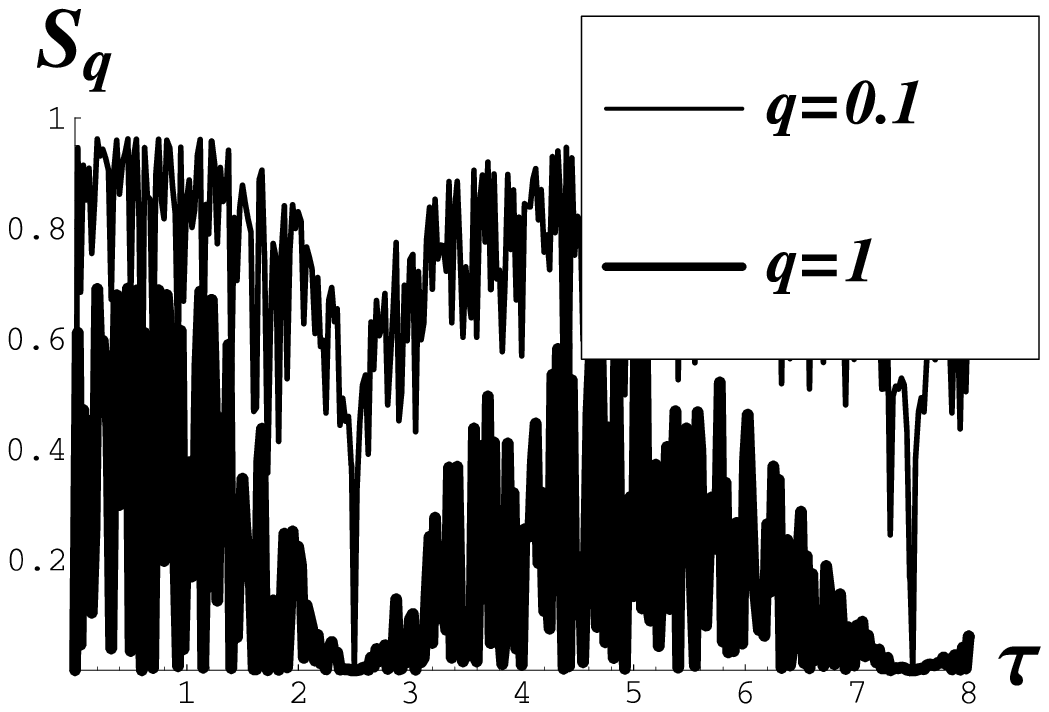}%
\epsfxsize=6.5truecm\epsfbox{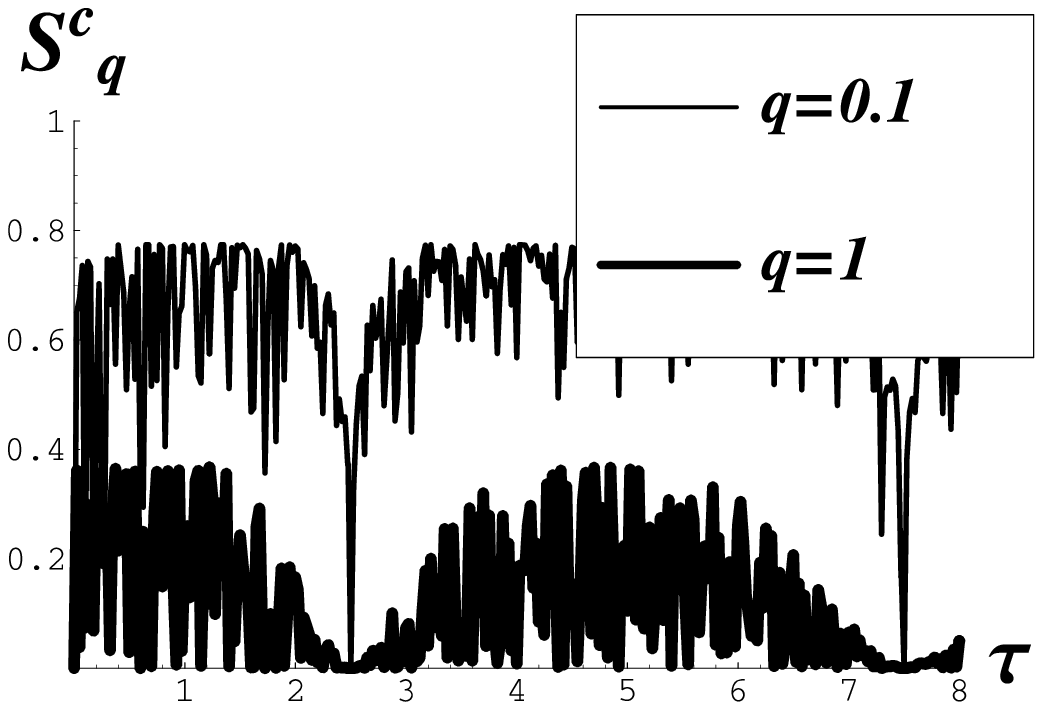}
\epsfxsize=6.5truecm\epsfbox{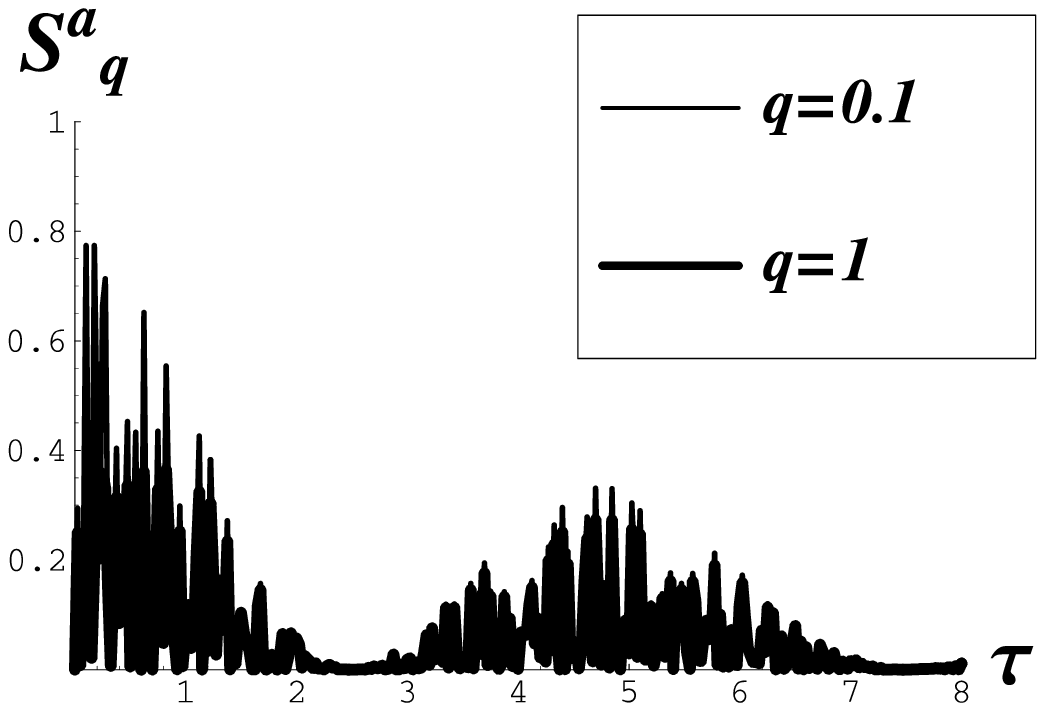}}

\vspace*{-8cm}\hspace{-1.5cm}\centerline{\epsfxsize=6.5truecm\epsfbox{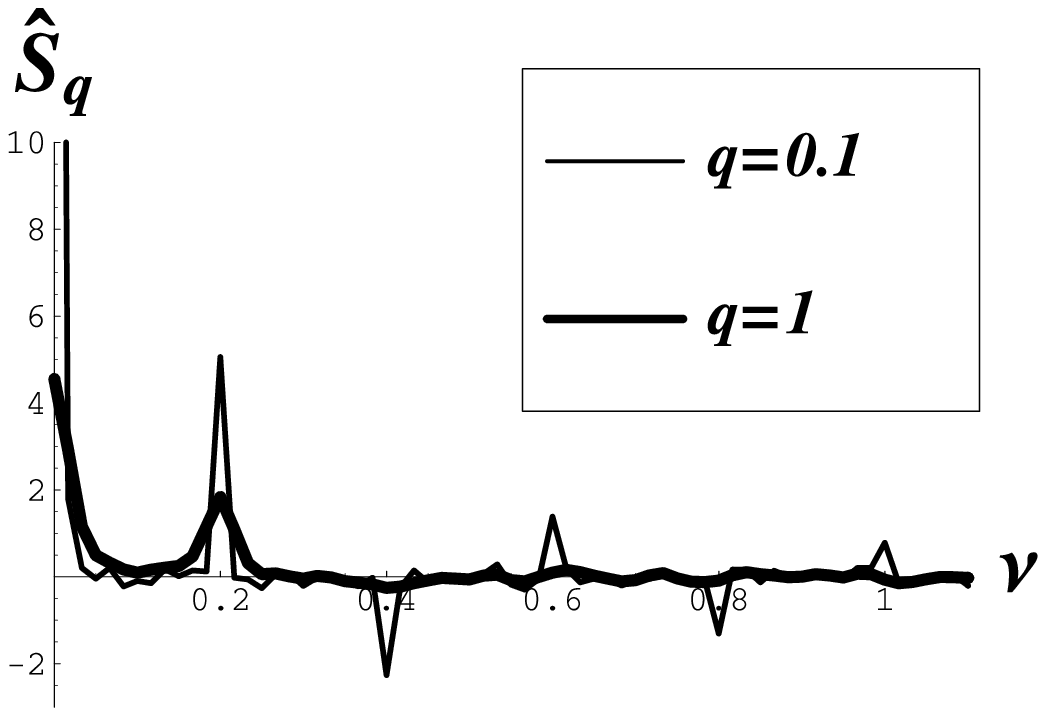}%
\epsfxsize=6.5truecm\epsfbox{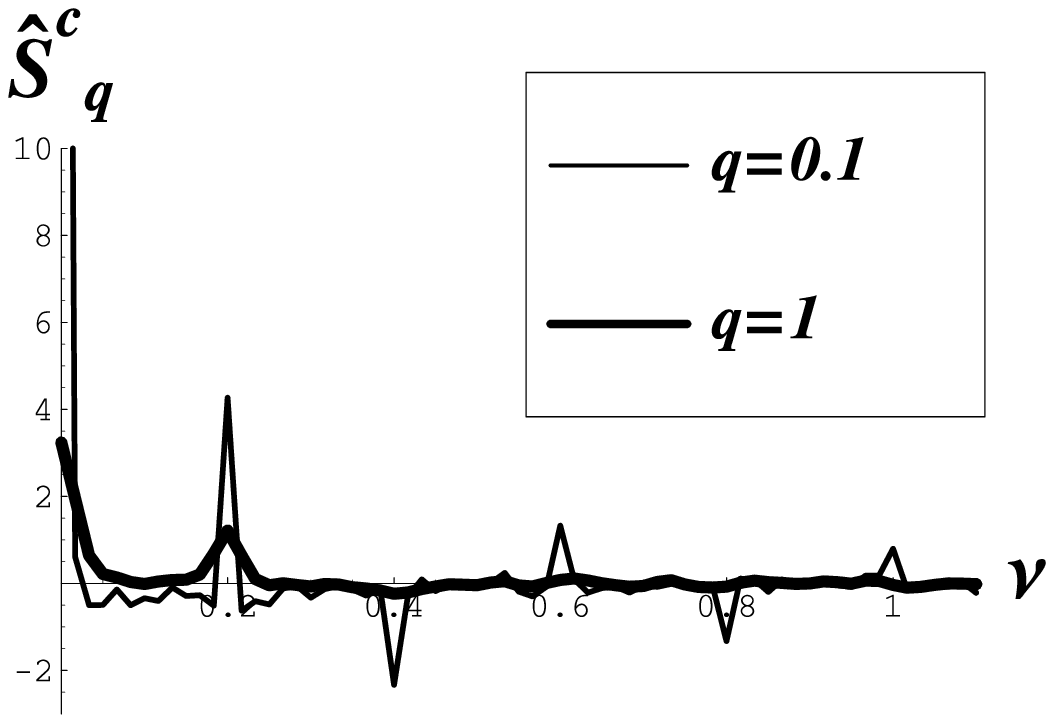}%
\epsfxsize=6.5truecm\epsfbox{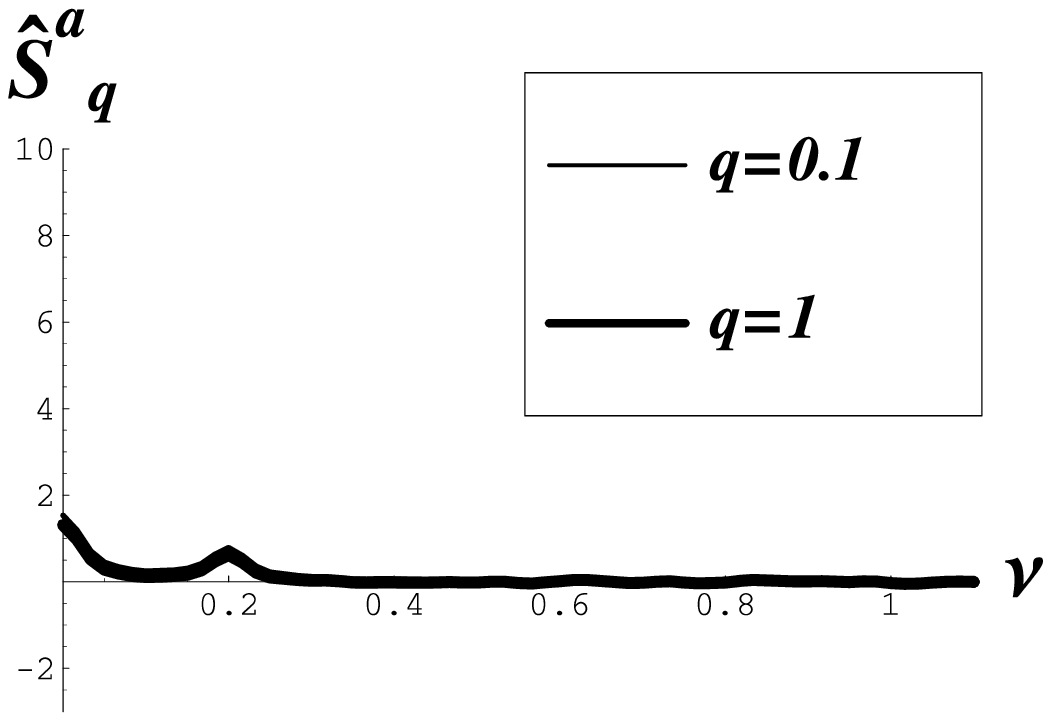}}

\vspace*{-8cm}\caption{The plots of entropies \Ref{QS} for $q=0.1$
and \Ref{Shannon} ($q=1$) and their spectra for $\nu' = 0.1,
c=0.1$ and for Brownian oscillator with noise. Comparison of
dynamic Shannon ($q=1$) and Tsallis ($q=0.1$) entropies shows the
amplification of DTE for small values of TCF. One can conclude
that DTE acts as a magnifier for small values of TCF.}
\label{Toy3Nois}
\end{figure}

The real signal from alive systems often contains noise (see Ref.
\cite{YulEmeGaf04}). For this reason we suggest the model of
Brownian oscillator with noise. We consider the following model of
TCF:
\be
M_0(t) = R(t)\cos (2\pi\nu' t) e^{-c|t|},\label{OscillatorNois}
\ee
where $R(t)$ denotes the random numbers in interval $(-1,1)$ and
$R(0)=1$. Therefore the function $R(t)$ makes random the amplitude
of oscillation but this is not the case for frequency and damping
parameters. At the beginning we know the frequency of
oscillations. The TCF has more complicate form which is more close
to real dates. The random numbers describe a noise which usually
appears in an experiment. The time dependence of entropy is much
more complicate but nevertheless in this case the DTE works
better. We observe the appearing peaks in places, which we know
the peaks must be, but they disappeared in noise. There is another
observation: the noise is better for frequency spectrum. The peaks
for small frequencies looks better (see Fig. \ref{Toy3Nois}).

\section{Density fluctuations in hydrodynamical limit}\label{Sec:5}

The TCF of density fluctuations in hydrodynamical limit was
calculated by Landau and Placzek \cite{ResDeL77}. It describes
scattering of light in liquid in hydrodynamics limit when $k\to
0$. The TCF has the following form
\bd
M_0(t) = \alpha e^{-\gamma' k^2 t}\cos \vartheta_s k t + (1 -
\alpha) e^{-\sigma' k^2 t},
\ed
where $\gamma' = \fr{1}{2\rho}\left(\fr 43 \eta + \zeta + \kappa
\left[\fr 1{c_v} - \fr 1{c_p}\right]\right),\ \sigma' =
\fr{\kappa}{\rho c_p},\ \alpha = c_v/c_p$. Here the
$c_v,c_p,\kappa, \eta, \zeta, \linebreak[1] \rho, \linebreak[1]
\vartheta_s$ are specific heat capacities in units of mass at
constant volume and constant pressure, the coefficient of thermal
conductivity, the coefficient of shear viscosity and volume
viscosity, the mass density and sound velocity, correspondingly.

The spectrum of this TCF contains three peaks. The central
Rayleigh peak at zero frequency describes isothermal propagation
of sound. Two symmetric peaks at frequencies $\omega = \pm
\vartheta_s k$ describe adiabatic propagation of sound with
damping (Brillouin doublet).

It is more suitable to define dimensionless time $\tau$ by
relation $\tau = \vartheta_s k t/2\pi$. Then the position of
Brillouin doublet will be at the dimensionless frequency $\nu =
\omega/2\pi = 1$, and TCF will take the following form
\bd
M_0(\tau) = \alpha e^{-\gamma k 2\pi\tau}\cos 2\pi\tau + (1 -
\alpha) e^{-\sigma k 2\pi\tau},
\ed
where $\gamma = \fr{1}{2\rho \vartheta_s}\left(\fr 43 \eta + \zeta
+ \kappa \left[\fr 1{c_v} - \fr 1{c_p}\right]\right),\ \sigma =
\fr{\kappa}{\rho c_p \vartheta_s},\ \alpha = c_v/c_p$. We consider
a specific medium -- Helium at temperature $T = 20^o C$ and
pressure $p=1 b$. In this case we have \cite{Vargaftik}: $\alpha
\approx 0.56,\vartheta_s \approx 272\ m/c, \gamma \approx 6\cdot
10^{-9}m^{-1}, \sigma \approx 7\cdot 10^{-9}m^{-1}$. We make
calculations for $k=2\cdot 10^{7}m^{-1}$.

\begin{figure}
\hspace{-1.5cm}\centerline{\epsfxsize=6.5truecm{\epsfbox{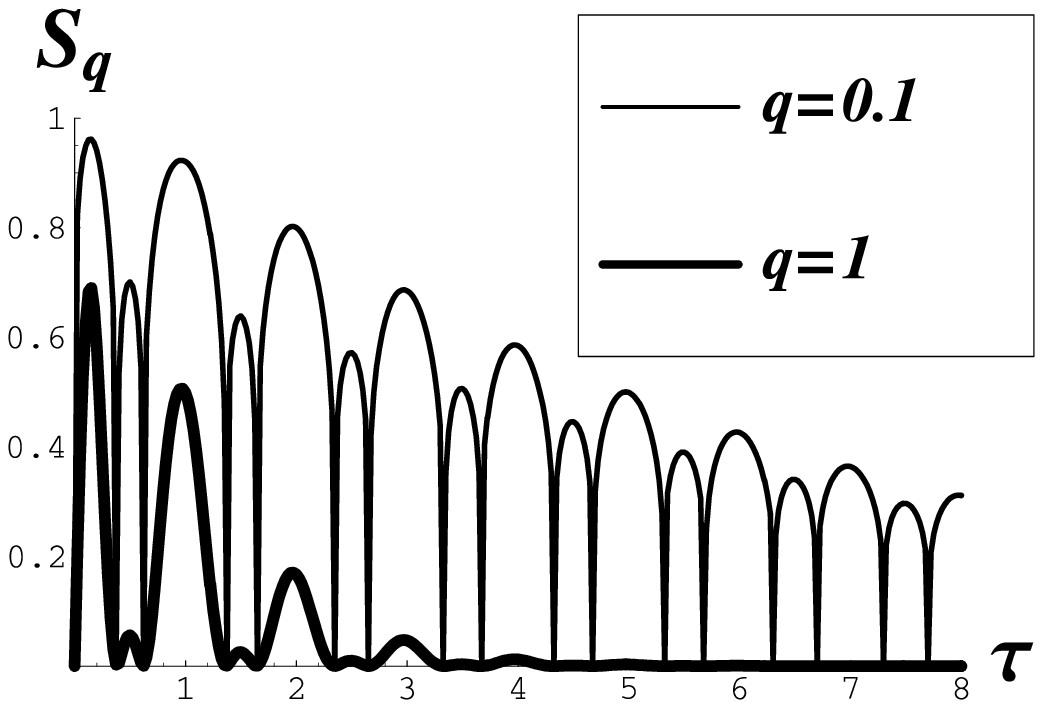}}%
\epsfxsize=6.5truecm\epsfbox{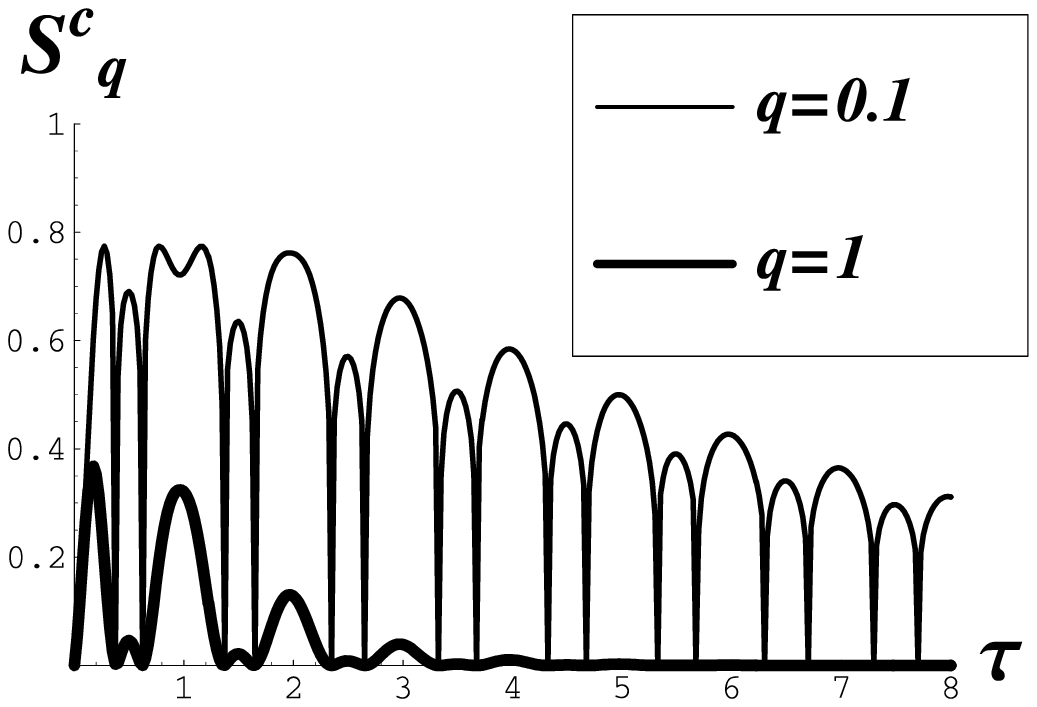}%
\epsfxsize=6.5truecm\epsfbox{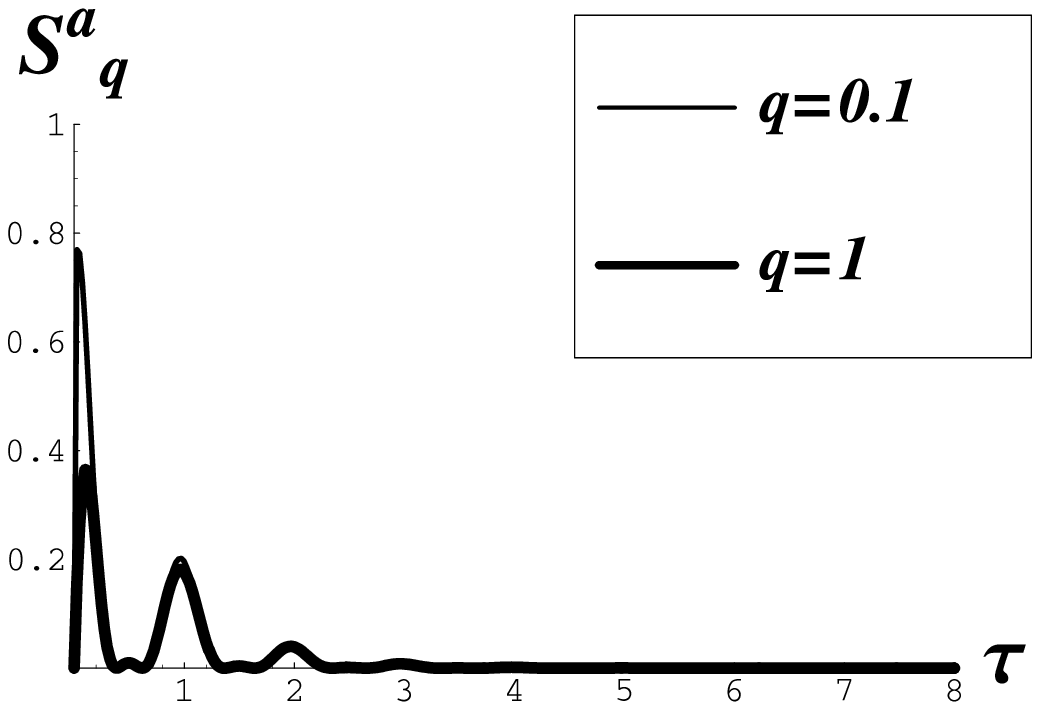}}

\vspace*{-8cm}\hspace{-1.5cm}\centerline{\epsfxsize=6.5truecm\epsfbox{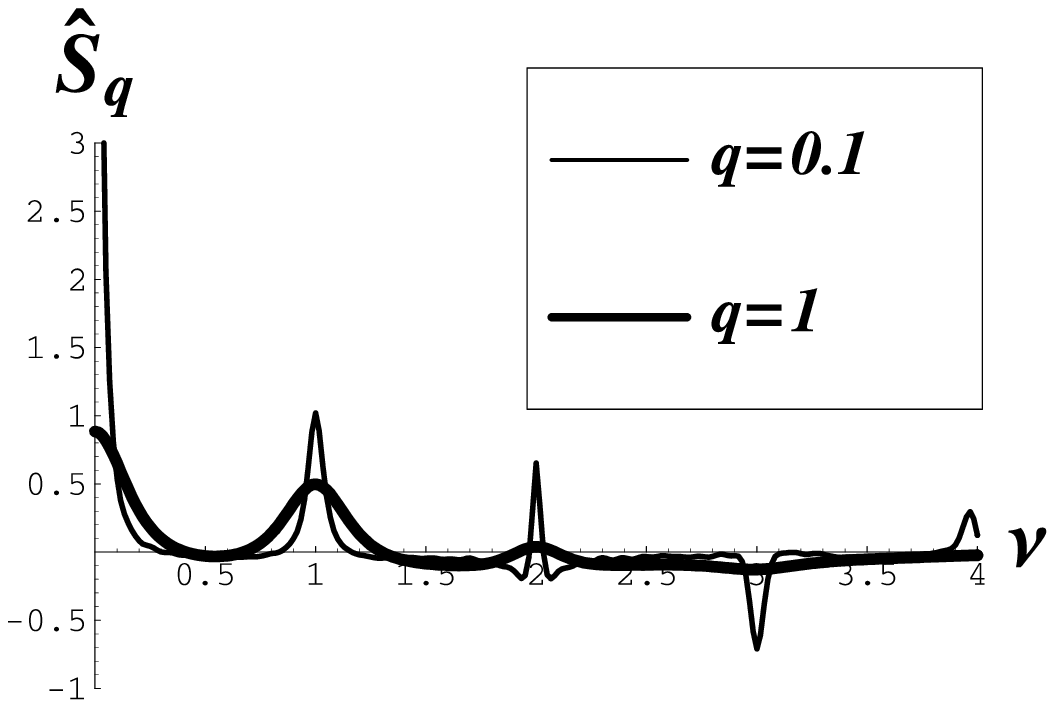}%
\epsfxsize=6.5truecm\epsfbox{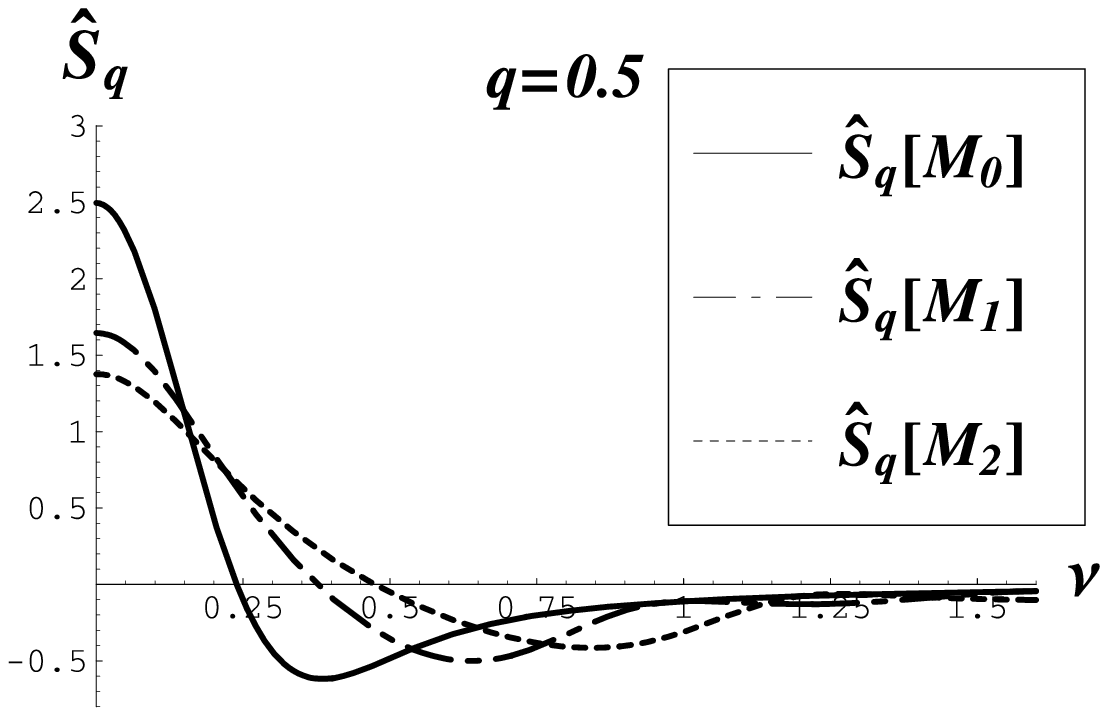}%
\epsfxsize=6.5truecm\epsfbox{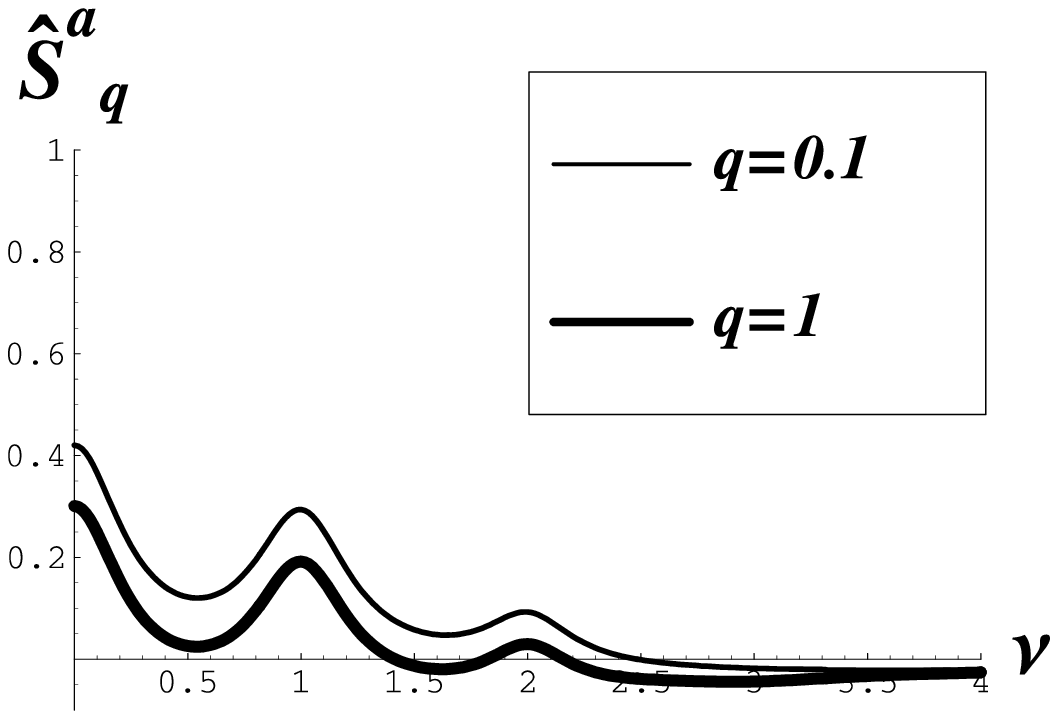}}

\vspace*{-8cm}\caption{Plots of entropies \Ref{QS} for $q=0.1$ and
\Ref{Shannon} ($q=1$) and their spectrum for Landau-Placzek TCF.
Here we use the following notations $S_q=S_q^0,\ S^c_q =
S_q^{0c},\ S^a_q = S_q^{0a}$, and $\widehat{S}_q=\widehat{S}_q^0,\
\widehat{S}^c_q=\widehat{S}_q^{0c},\ \widehat{S}^a_q
=\widehat{S}_q^{0a}$.}\label{lp}
\end{figure}

In the Fig.\ref{lp} we reproduce time and frequency dependencies
entropies defined before for two values of parameter $q=1,0.1$.
There is periodicity over $\tau$ with unit period which gives the
appearance specific peaks in the frequency spectrum (Brillouin
doublet). We observe the same picture as in previous section.
Decreasing the parameter of nonextensivity $q$ leads to increasing
small peaks in entropies $S_{qn}$ and $S_{qcm_n}$, whereas
quantity $S_{qam_n}$ is changed insufficiently. The frequency
spectrum entropies $S_{qn}$ and $S_{qcm_n}$ becomes more sharp.
The shape of frequency spectrum $S_{qam_n}$, in fact, does not
change.

The non-Markovity parameter of this system has been calculated
earlier in Refs. \cite{YulKhu94,KhuYul95}. It was shown that in
hydrodynamical limit $k\to 0$ the spectrum of non-Markovity
parameter has a form of alternating Markovian and non-Markovian
levels.

\section{Ideal Gas}\label{Sec:6}

Let us consider the Fourier transformation of the fluctuation of
the particle number density of a system
\bd
\delta\rho_k(t) = \fr 1V \sum_{l=1}^N \exp (i\mathbf{kr}_l) - \fr
NV \delta_{k,0}.
\ed
For this case the initial TCF is calculated exactly and it has the
following form \cite{ResDeL77}:
\bd
M_0(t) =
\fr{\langle\delta\rho_k(t)\delta\rho_k(0)^*\rangle}{\langle|\delta\rho_k(0)|^2\rangle}=
e^{-t^2/t^2_r},\ t^2_r = 2m /k^2T.
\ed

For this variable the all frequencies $\omega_0^{(n)}$ are equal
to zero. The main relaxation frequencies in Eq. \Ref{MainSystem}
have the simple form
\bd
\Omega_n^2 = n \Omega_1^2,\ \Omega_1^2 = k^2T/m.
\ed

Let us rescale time $t\to \tau = t\Omega_1$, and the Laplace
transformation parameter $s\to c = s/\Omega_1$, and the Laplace
images $\widetilde{M}_n(s) \to \widetilde{m}_n(s) = \Omega_1
\widetilde{M}_n(s)$. In this case the hierarchy \Ref{MainSystem}
has the following form
\bd
\widetilde{m}_{n+1}(c) = \fr 1{n+1} \left\{\fr
1{\widetilde{m}_n(c)} - c\right\}.
\ed

The Laplace image $\widetilde{m}_0(c)$ may be found in close form:
\be
\widetilde{m}_0(c) = e^{\fr{c^2}2}\sqrt{\fr \pi 2}\
\mathrm{Erfc}[\fr c{\sqrt{2}}],\label{m0}
\ee
where $\mathrm{Erfc}(x) = 1- \mathrm{Erf}(x)$ -- additional
probability integral.

\begin{figure}
\hspace{-1.5cm}\centerline{\epsfxsize=9truecm\epsfbox{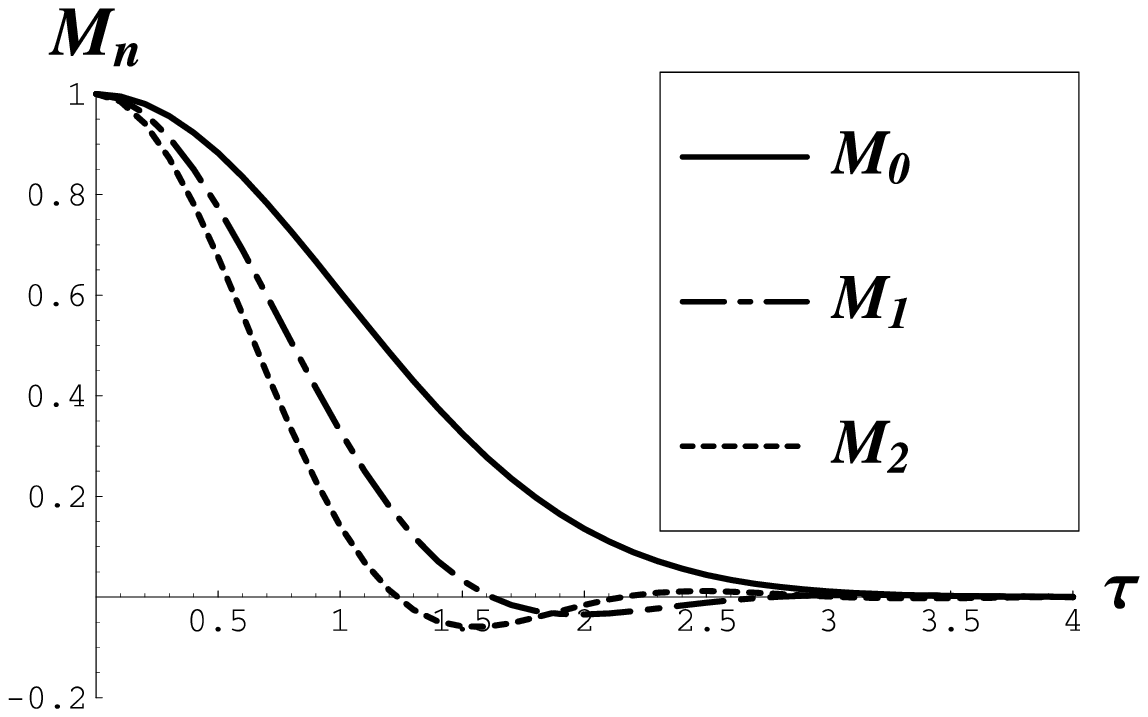}\epsfxsize=9truecm\epsfbox{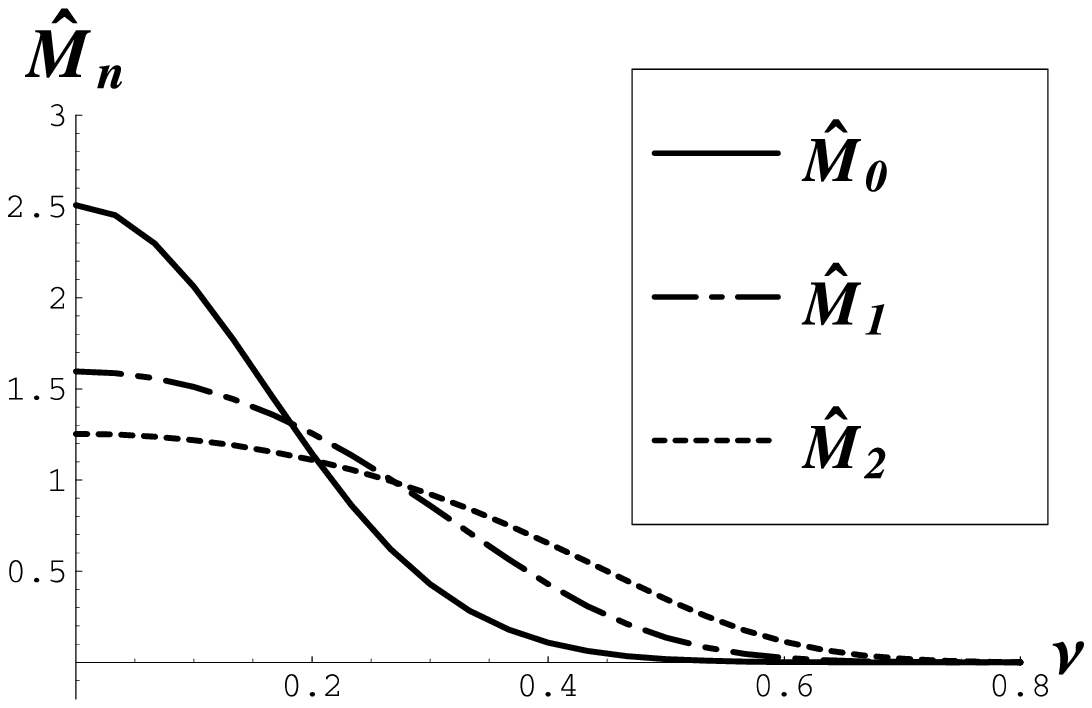}}

\vspace*{-11cm}\caption{The first normalized memory functions
$M_n(\tau)$ and their frequency spectrum $\widehat{M}_n(\nu)$ for
ideal gas.}\label{Fig1}
\end{figure}

\begin{figure}
\hspace{-1.5cm}\centerline{\epsfxsize=9truecm\epsfbox{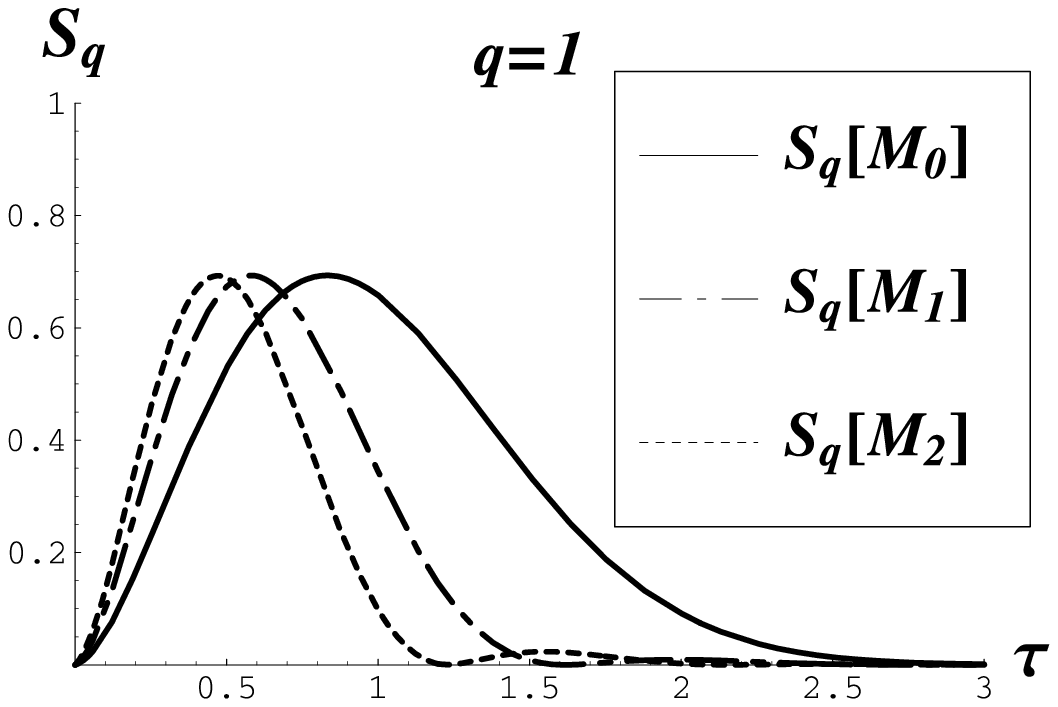}%
\epsfxsize=9truecm\epsfbox{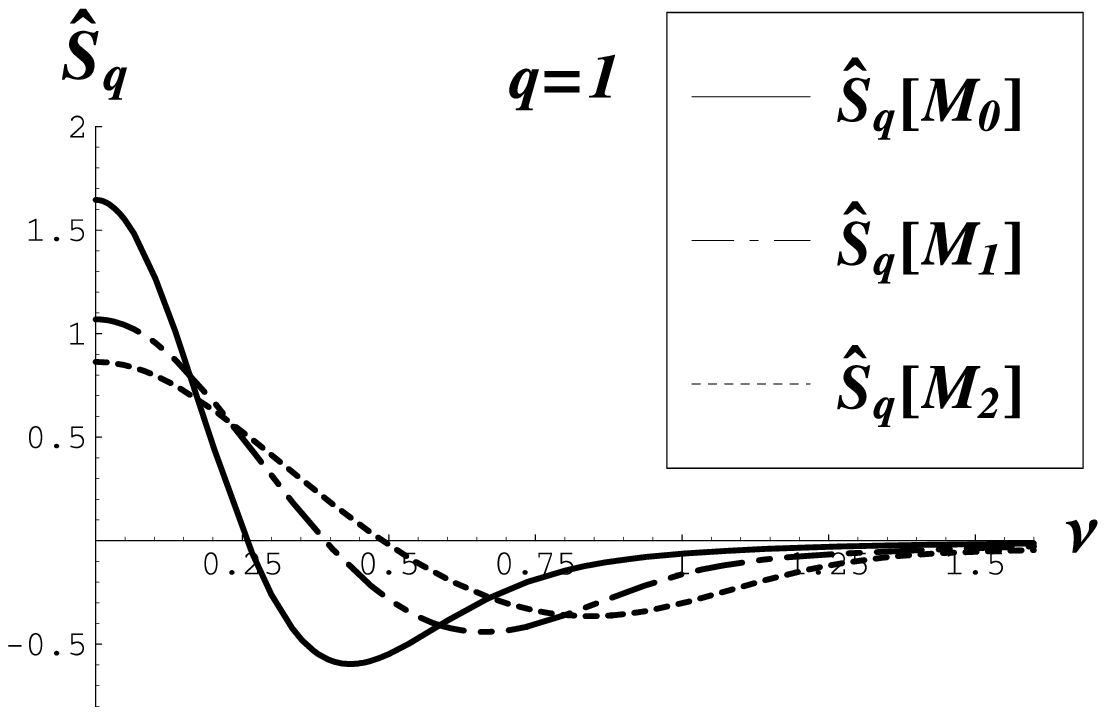}}

\vspace*{-11cm}\caption{The plot the time dependence of entropy
and its frequency spectrum for different relaxation levels
$n=0,1,2$ for ideal gas.}\label{Fig2}
\end{figure}

The inverse transformation may be represented in the following
form:
\bd
M_n(\tau) = \fr 1{2\pi i} \int_{\sigma - i\infty}^{\sigma +
i\infty}   e^{c\tau} \widetilde{m}_n(c) dc,
\ed
where $\sigma$ is greater then real part of zeros of
$\widetilde{m}_n(c)$. By using the expression \Ref{m0} we may set
$\sigma =0$ and by changing $c\to ix$ we obtain
\bd
M_n(\tau) = \fr 1{2\pi} \int_{-\infty}^{+\infty} e^{icx}
\widetilde{m}_n(ix) dx,
\ed
or in manifest form
\bnn
M_1(\tau) &=& \fr 1{2\pi} \int_{-\infty}^{+\infty} e^{i\tau x}
\left[\fr 1{\widetilde{m}_0(ix)} - ix\right] dx,\adb \\
M_2(\tau) &=& \fr 1{4\pi} \int_{-\infty}^{+\infty} e^{i\tau x}
\left[ \fr 1{\fr 1{\widetilde{m}_0(ix)} - ix} - ix\right] dx.
\enn
These formulas we will calculate numerically. The first three
memory functions and their spectrum are plotted in Fig.\ref{Fig1}.
In Fig. \ref{Fig2} we reproduce the plot of the entropy $S^n$ and
their frequency spectrum for $n=0,1,2$.

It is not difficult to show that the entropy $S^n$ \Ref{Shannon}
amounts to maximum value, $\ln 2$, at the point $|M_n|^2 = \fr
12$. For $n=0$ the position of maximum is at the point $\tau =
\sqrt{\ln 2} \approx 0.832$. The greater $n$, the smaller time of
maximum. The relaxation times $\tau_0 \approx 1.6453,\ \tau_1 =
1.0688,\ \tau_2 = 0.8630$. The parameter of non-Markovity,
\cite{ShuYul90,ShuYul91}
\bd
\epsilon_n = \tau_n/\tau_{n+1}
\ed
has the following values $\epsilon_0 \approx 1.54,\ \epsilon_1
\approx 1.24$. Note that these values very close to that
calculated in paper \cite{YulKhu94} directly for TCFs, $\epsilon_0
\approx 1.57,\ \epsilon_1 \approx 1.27$.

\begin{figure}[ht]
\hspace{-1.5cm}\centerline{\epsfxsize=9truecm\epsfbox{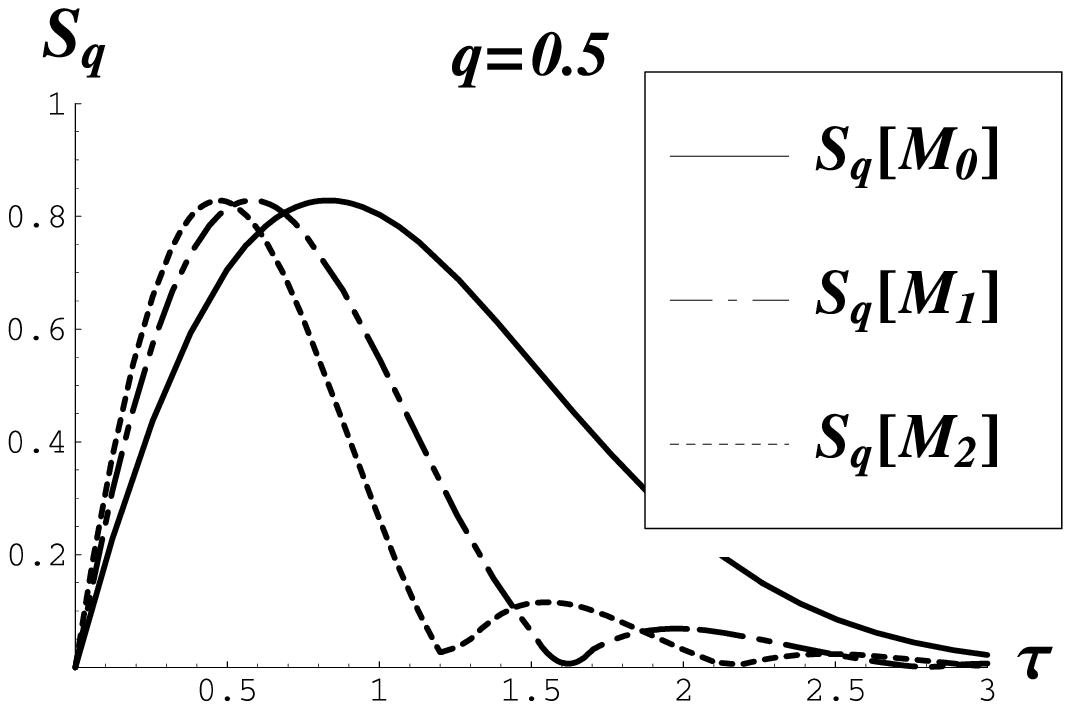}%
\epsfxsize=9truecm\epsfbox{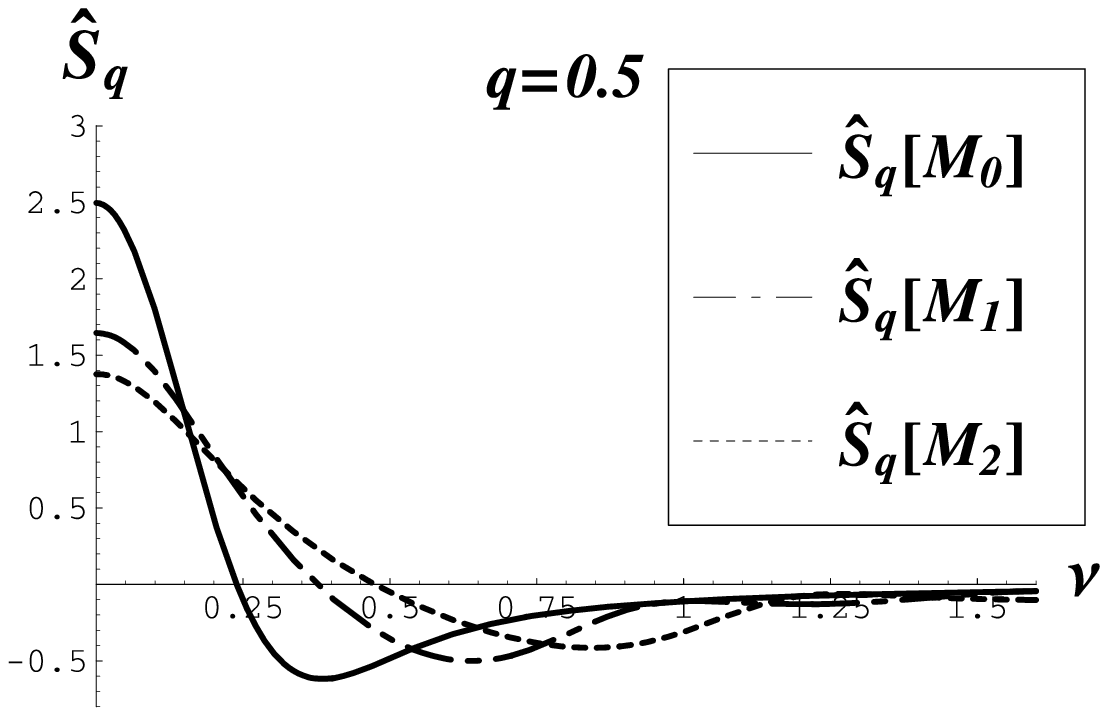}}

\vspace*{-11cm}\hspace{-1.5cm}\centerline{\epsfxsize=9truecm\epsfbox{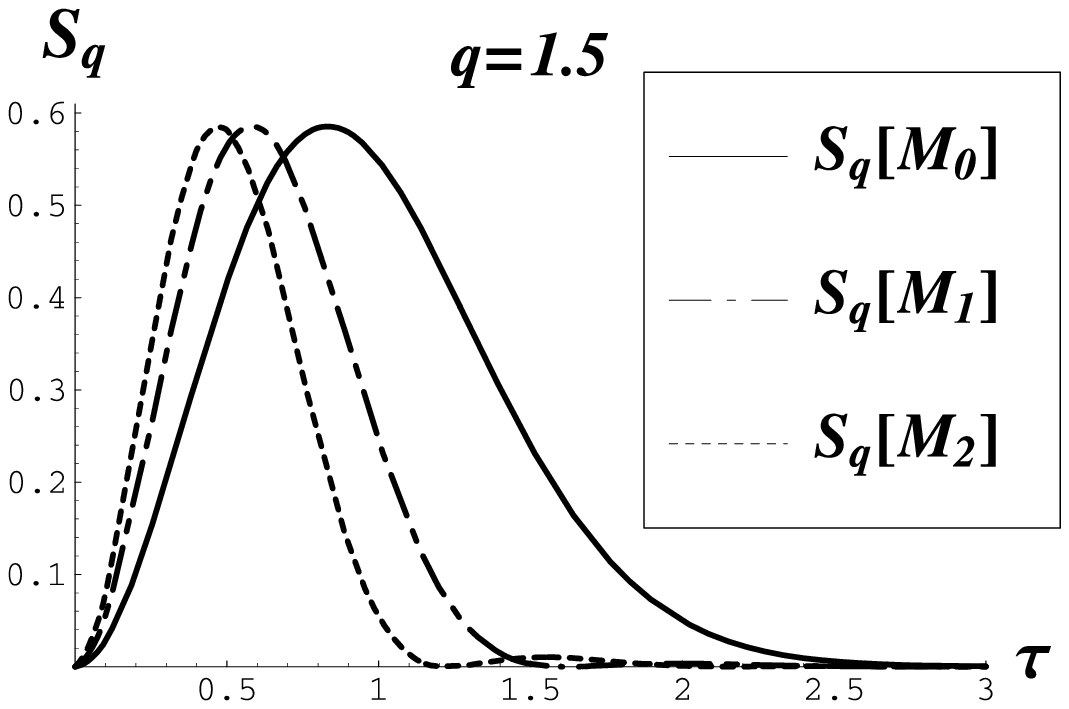}%
\epsfxsize=9truecm\epsfbox{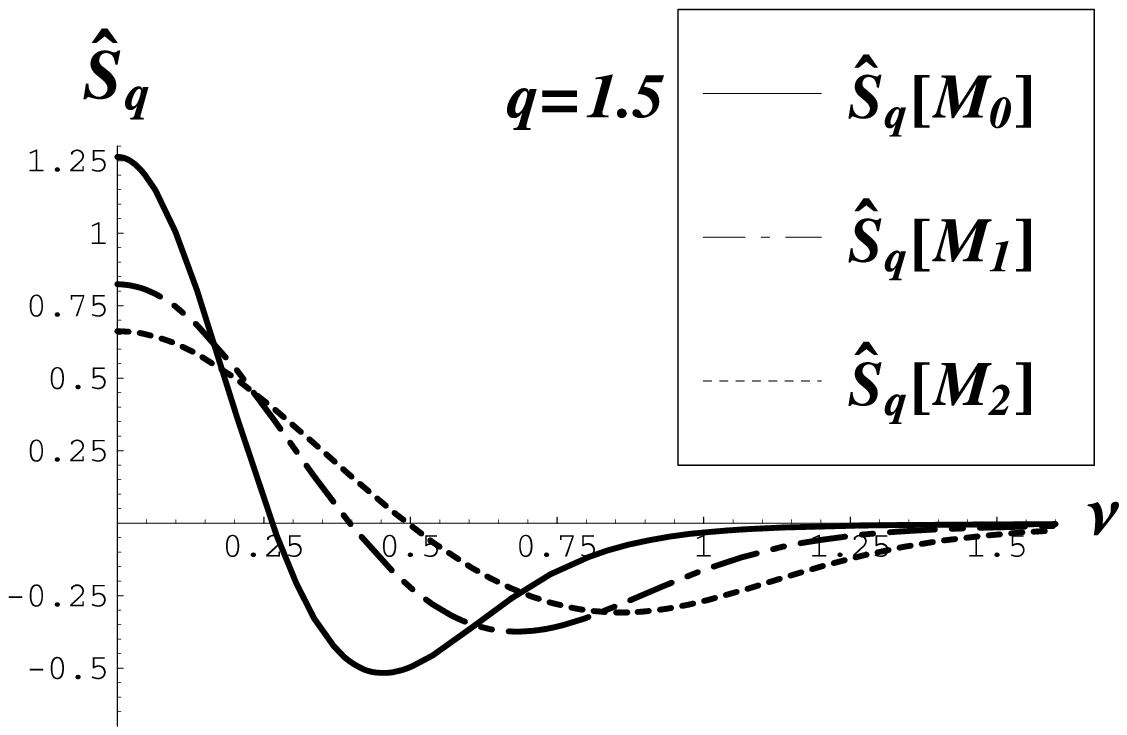}}

\vspace*{-11cm}\caption{The plots of time dependent entropy
$S^n_q$ and their frequency spectrums for $q=1/2$ and
$q=3/2$.}\label{Fig3}
\end{figure}
\begin{figure}
\hspace{-1.5cm}\centerline{\epsfxsize=9truecm\epsfbox{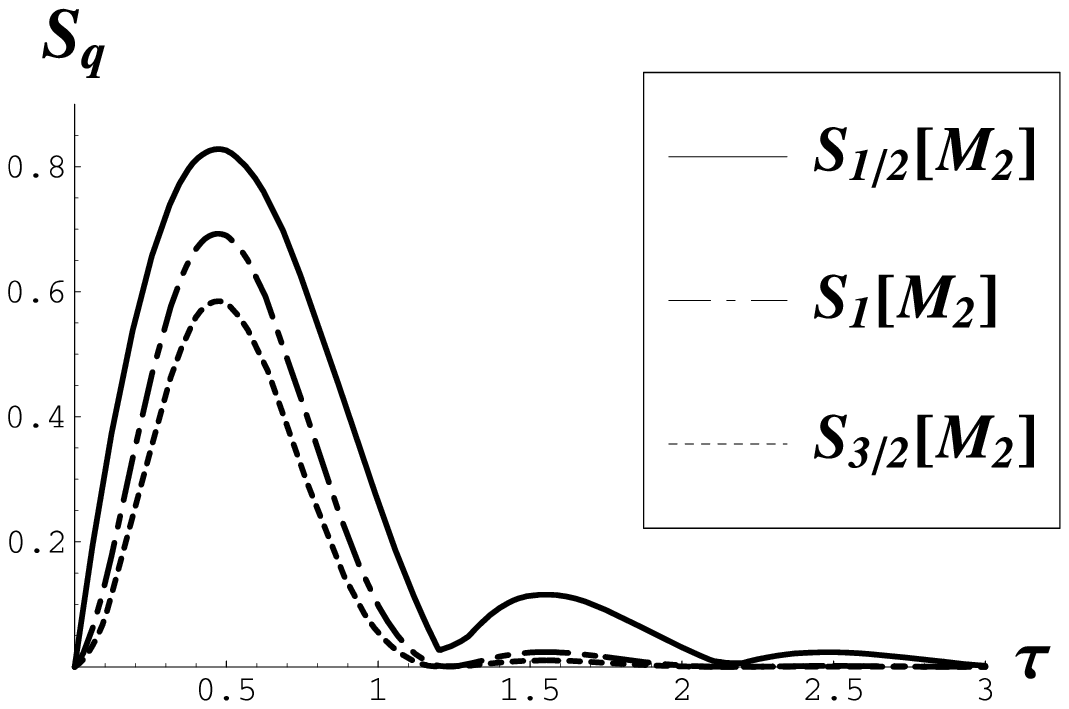}%
\epsfxsize=9truecm\epsfbox{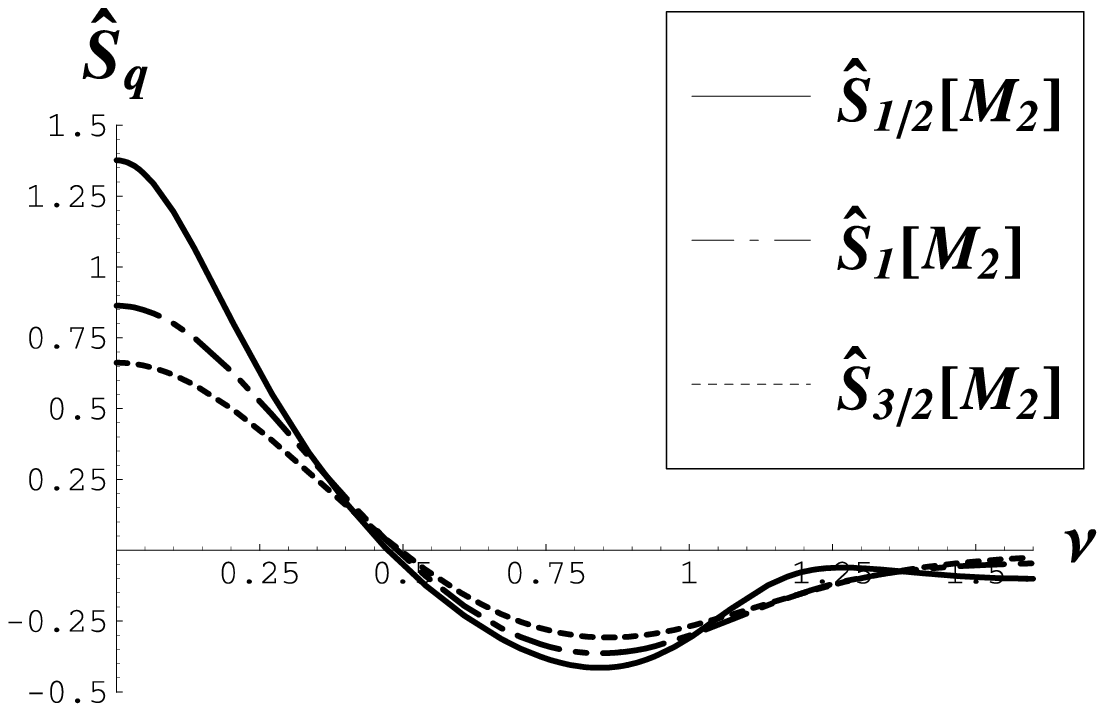}}

\vspace*{-11cm}\caption{The time dependent entropy $S^n_q$ and
their frequency spectrum for $n=2$ and for
$q=0.5,1,1.5$.}\label{Fig4}
\end{figure}
\begin{figure}
\hspace{-1.5cm}\centerline{\epsfxsize=9truecm\epsfbox{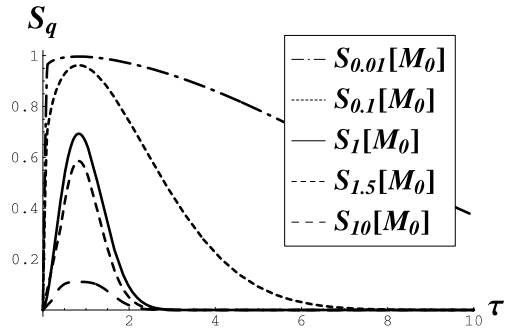}}

\vspace*{-11cm}\caption{The time dependence of the entropy $S^n_q$
and its frequency spectrum for six values  $q=0.001, 0.1, 0.5, 1,
1.5, 10$ and for $n=0$. One observe that the increasing Tsallis'
parameter $q$ leads to trample a quantity of $S_q$ in domain of
short time. Due to this fact one can make sufficient amplification
in domain of short time owing to variation parameter
$q$.}\label{Fig5}
\end{figure}

In the Fig. \Ref{Fig3} we give the plots of entropy $S^n_q$ and
their spectrums for $q=1/2$ and $q=3/2$. In order to show what
happens if we will vary the value of $q$ we reproduce in Fig.
\Ref{Fig4} the entropy $S^n_q$ and its spectrum for $n=2$ and for
$q=0.5,1,1.5$. To show the dependence $S^n_q$ for more wide range
of $q$ in Fig. \Ref{Fig5} we give entropies for more wide range of
$q$.

By using these three quantities we define seven different spectra
of non-Markovity parameter:
\bs
\bn
\epsilon_{qn} &=& \fr{\tau_{qn}}{\tau_{qn+1}},\ \epsilon_{qc_n} =
\fr{\tau_{qcn}}{\tau_{qcn+1}},\ \epsilon_{qa_n} =
\fr{\tau_{qan}}{\tau_{qan+1}},\adb\\
\epsilon_{qca_n} &=& \fr{\tau_{qcn}}{\tau_{qan+1}},\
\epsilon_{qac_n} = \fr{\tau_{qan}}{\tau_{qcn+1}},\ \epsilon_{qcn}
= \fr{\tau_{qcn}}{\tau_{qn+1}},\ \epsilon_{qan} =
\fr{\tau_{qan}}{\tau_{qn+1}}.\label{Epca}
\en
\es

\begin{figure}
\centerline{\epsfxsize=9truecm\epsfbox{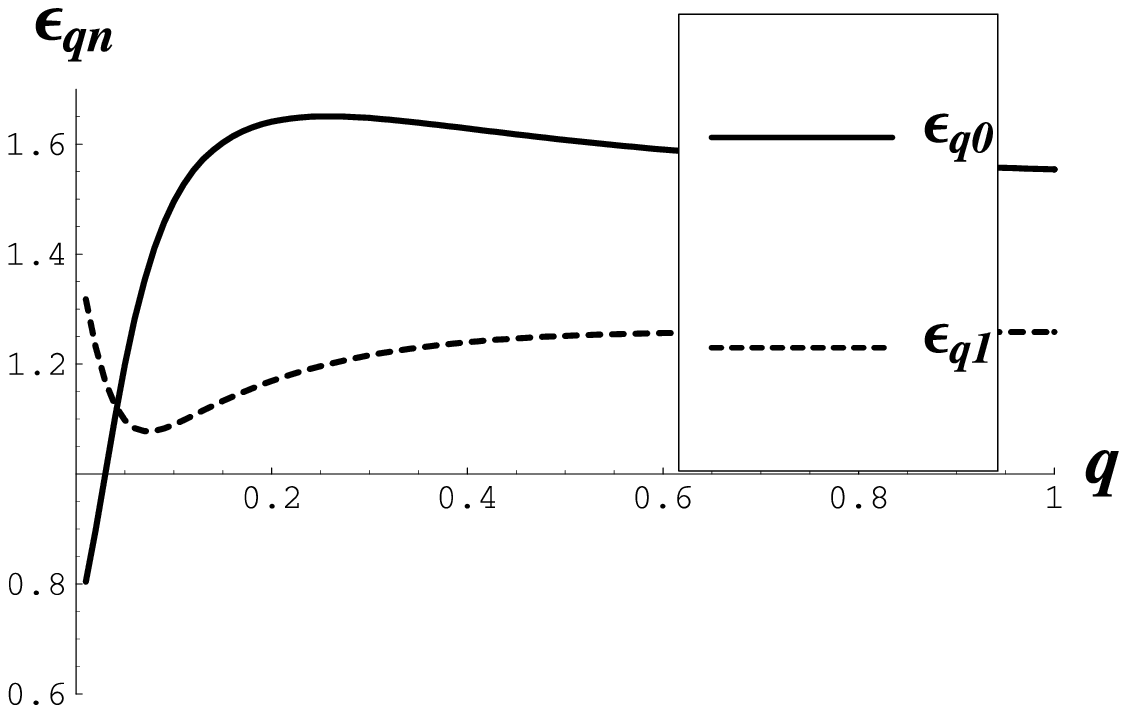}}

\vspace*{-11cm}\hspace{-1.5cm}\centerline{\epsfxsize=9truecm\epsfbox{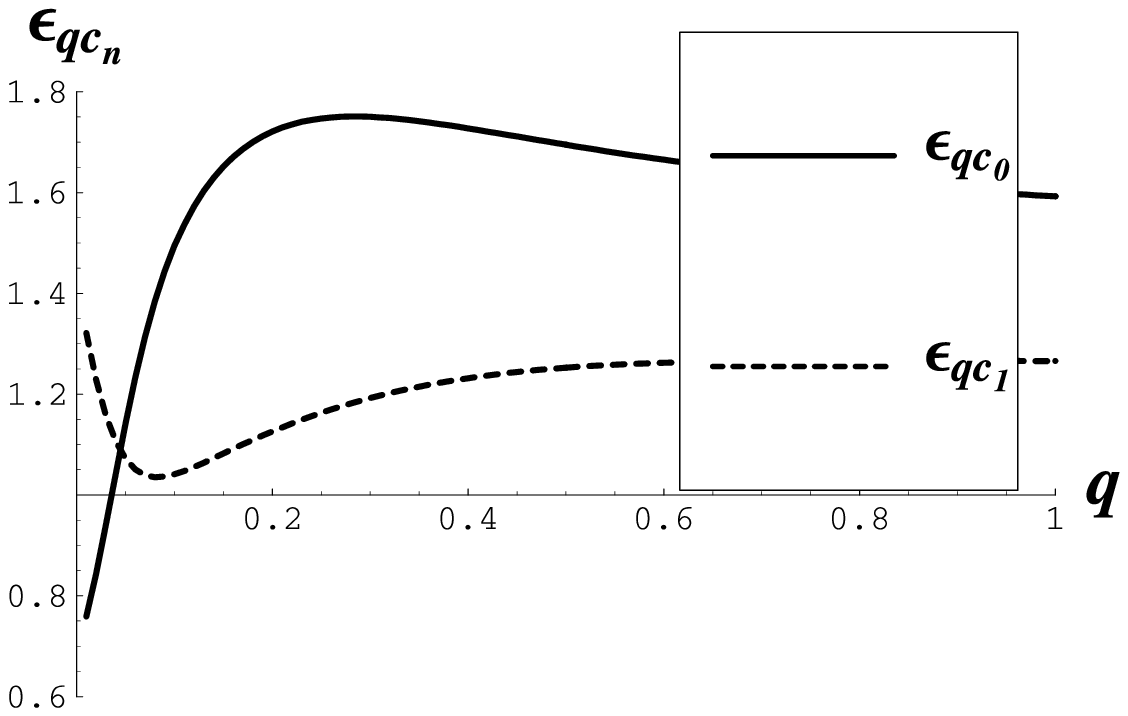}%
\epsfxsize=9truecm\epsfbox{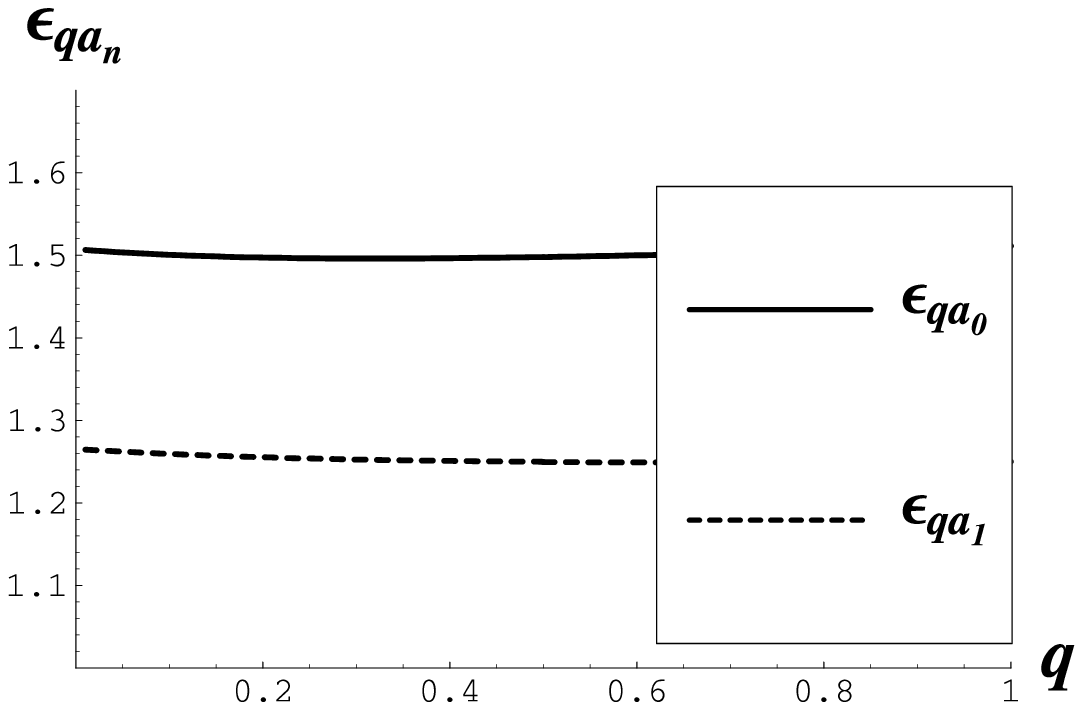}}

\vspace*{-11cm}\hspace{-1.5cm}\centerline{\epsfxsize=9truecm\epsfbox{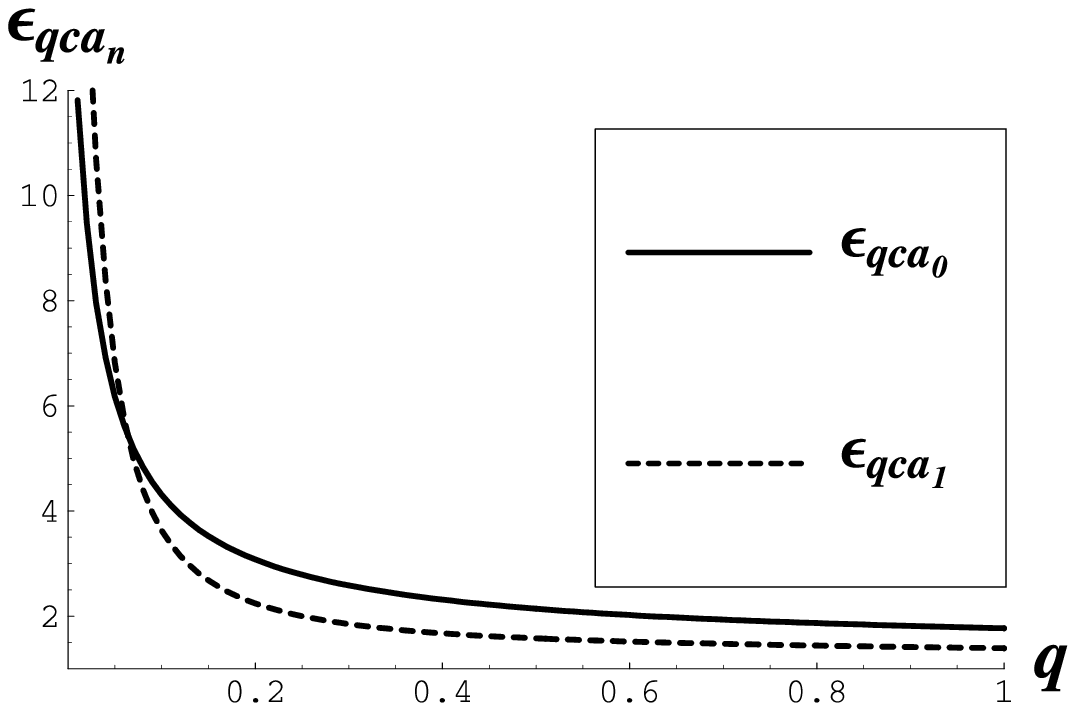}%
\epsfxsize=9truecm\epsfbox{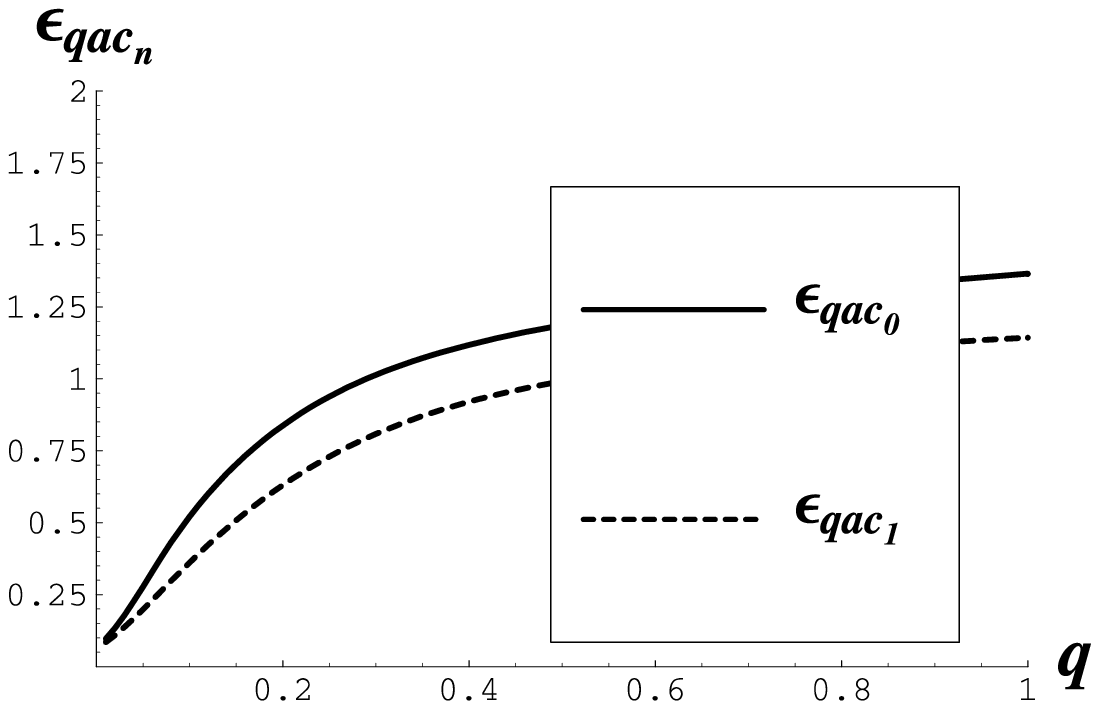}}

\vspace*{-11cm}\hspace{-1.5cm}\centerline{\epsfxsize=9truecm\epsfbox{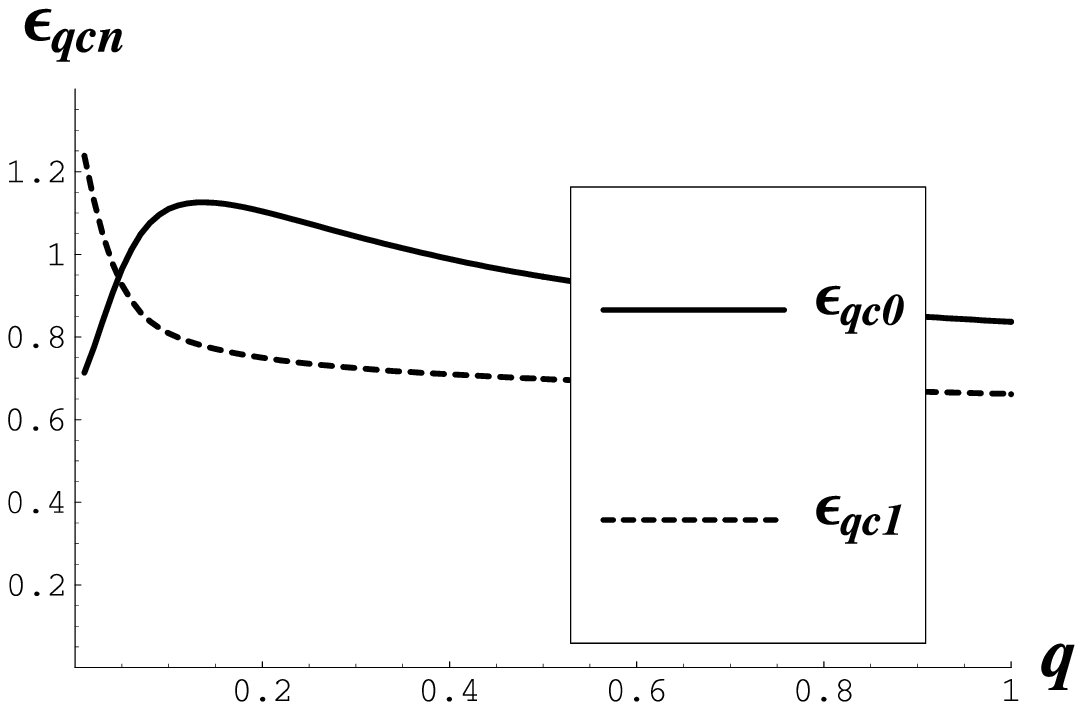}%
\epsfxsize=9truecm\epsfbox{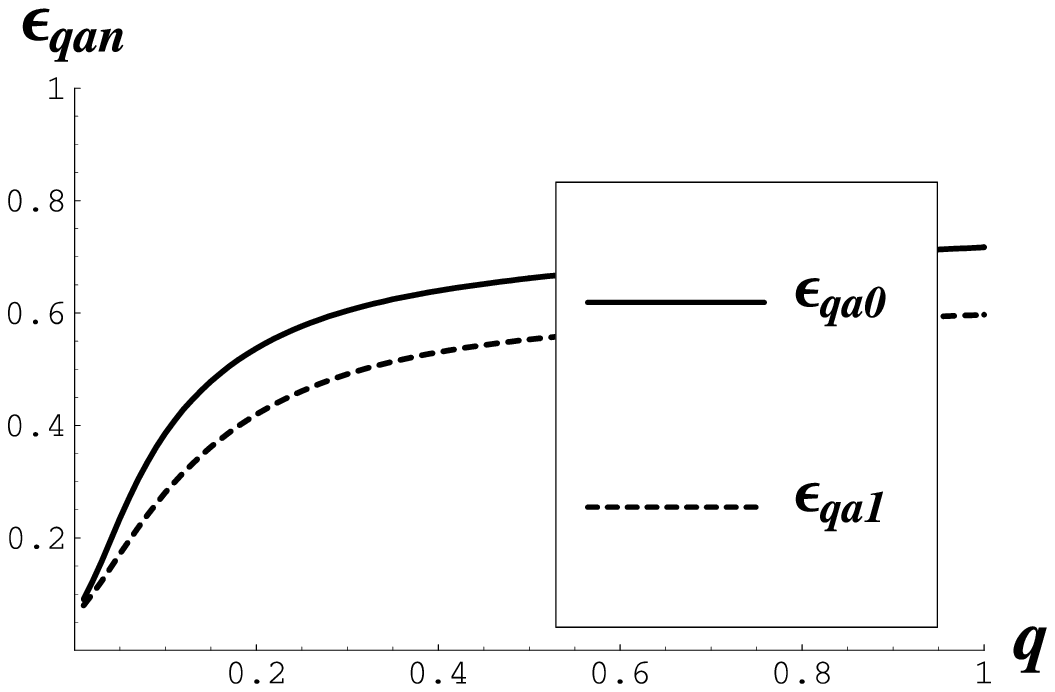}}

\vspace*{-11cm}\caption{The spectrum of all parameters of
non-Markovity as function of $q$. The influence of parameter $q$
is most effectively for small values of $q\ll 1$. In all cases we
observe strong non-Markovity and clearly marked effect of
statistical memory. The only case where we may sufficiently
increase the non-Markovity properties by decreasing parameter
$q\to 0$ is cross term $ca$ (see Eq. \Ref{Epca}). For small enough
$q\ll 1$ the non-Markovity parameter $\epsilon_{qca_n}$ may reach
great value $\epsilon_{qca_n} \gg 1$.}\label{Fig7}
\end{figure}

The plots of all quantities are shown in Fig. \Ref{Fig7}. As
expected at the beginning more interesting situations are possible
for small values of $q$. For great value of $q$ all lines tend to
constants.

\section{Conclusion}\label{Sec:7}

Let us summarize our results. We considered the dynamic Tsallis
entropy and applied it for model systems. In accordance of Refs.
\cite{YulKle98,YulGaf99,YulGafYulEme02} the square of TCF is
regarded as probability of dynamic state. We used the time
dependent entropies as well as their frequency spectra. We
considered four model TCFs which, nevertheless, have physical
sense. The first and second models describe the  motion of
Brownian oscillator without and with noise. The noise is modelled
by generator of random numbers. Third model is the Landau-Placzek
model for TCF of particle density fluctuations in hydrodynamic
limit. The last model is ideal gas, the relaxation of particle
density fluctuations.

In all model we considered different values of parameter of
nonextensivity $q$. All entropies have the same structure zeros
and extremums as square of TCF. The magnitude of extremums
sufficiently depends on value of $q$. Small values of $q$ work as
non-linear magnifier: the smaller magnitude of entropy the greater
magnification. Great values of $q$ do in opposite way as
demagnified lens. Concerning the frequency spectra of entropies we
observe that for small values of $q$ the peaks become more sharp
and more large. For systems with noise this property works better.
It is possible to reveal peaks even if they are lost in noise.

For ideal gas we additionally calculated the spectrum of parameter
non-Markovity by using different definition of relaxation time. By
using three kind of information Tsallis entropy we defined seven
kind of spectrum of parameter non-Markovity. We observe that all
of these parameters (except $\epsilon_{qca_n}$) are close to unit.
It means that the ideal gas remains non-Markovian system for
arbitrary value of $q$. It is in qualitative agreement with Ref.
\cite{Abe99}. The variation $q$ from unit does not mean the
appearance of new interactions. The system is became ideal and
non-Markovian from this point of view.

Our analysis allows us to conclude, that use of dynamic Tsallis's
entropy extends essentially possibilities of the stochastic
description of model physical systems. Advantage of application of
DTE is that it allows to strengthen or suppress fluctuations
either in low-frequency, or in high-frequency areas of a spectrum.
Similar supervision opens appreciable prospects in the field of
the study of real complex systems of wildlife where dynamic states
of physiological and pathological systems are very important.

\begin{acknowledgements}
This work was supported in part by the Russian Foundation for
Basic Research Grants N 02-02-16146, 03-02-96250 and by the
Russian Humanitarian Scientific Foundation Grant N 03-06-00218a.
\end{acknowledgements}


\begin{thebibliography}{10}
\bibitem{Abe99} Abe S 1999 \textit{Physica A} \textbf{269} 403
\bibitem{AczDar75} Acz\' el A., Dar\' oczy Z. 1975 \textit{On mesures of information
and their characterizations}  (Academic Press, New York)
\bibitem{BecSch93} Beck C and Schlogl F 1993 \textit{Thermodynamics of Chaotic
Systems} (Cambridge University Press, Cambridge)
\bibitem{GoyHan00} Goychuk I and Hanggi P 2000 \textit{Phys. Rev.}
\textbf{E61} 4272
\bibitem{KatHas95} Katok A., Hasselblatt B. 1995 \textit{Introduction
to the Modern Theory of Dynamical Systems} (Cambridge University
Press, Cambridge)
\bibitem{Kau93} Kaufman S A 1993 \textit{The Origins of Order --
Self-Organization and Selection in Evolution} (Oxford University
Press, Oxford)
\bibitem{KhuYul95} Khusnutdinov N R and Yulmetyev R M 1995
\textit{Teor. i Matem. Fiz.} \textbf{105} 292 (Engl. Transl. 1995
\textit{Theor. Math. Phys.} \textbf{105} 1426)
\bibitem{Kli98} Klimontovich Yu L 1998 \textit{Phys. Scr.}
\textbf{58} 549
\bibitem{LatBarRapTsa00} Latora V, Baranger M, Rapisarda A and Tsallis C
2000 \textit{Phys. Lett.} \textbf{A273} 97
\bibitem{Mor65} Mori H 1965 \textit{Prog. Theor. Phys.}
\textbf{33} 423; \textbf{34} 765
\bibitem{ReeSim72} Reed M and Simon B 1972 \textit{Methods of Modern
Mathematical Physics} (New York: Academic)
\bibitem{ResDeL77} R\'esibois P and De Leener M 1977
\textit{Classical kinetic theory of fluids} (New York: Wiley,
1977)
\bibitem{ScaSeaFer72} Scapino D J, Sears M and Ferel R A 1972 \textit{Phys. Rev.}
\textbf{B6} 3409
\bibitem{ShuYul90} Shurygin V Yu, Yulmetyev R M, and Vorobjev V V 1990 \textit{Phys.
Lett.} \textbf{148A} 199
\bibitem{ShuYul91} Shurygin V Yu and Yulmetyev R M 1991
\textit{Zh. Eksp. Teor. Fiz.} \textbf{99} 144 (Engl. Transl. 1991
\textit{Sov. Phys.-JETP} \textbf{72} 80)
\bibitem{Tsa88} Tsallis C 1988 \textit{J. Stat. Phys.}
\textbf{52} 479
\bibitem{Tsa99} Tsallis C 1999 \textit{Braz. J. Phys.}
\textbf{29} 1
\bibitem{Vargaftik} Vargaftik N B \textit{Handbook of Physical
Properties of Liquids and Gases: Pure Substances and Mixtures}
(Hemisphere Pub; 2nd edition 1983)
\bibitem{YulKhu94} Yulmetyev R M and Khusnutdinov N R 1994
\textit{J. Phys.} \textbf{A27} 5363
\bibitem{YulKle98} Yulmetyev R M and Kleiner M Ya 1998
\textit{Nonlinear Phenomena in Complex Systems} \textbf{1} 80
\bibitem{YulGaf99}  Yulmetyev R M and Gafarov F M 1999
\textit{Physica} \textbf{A273} 416\\
Yulmetyev R M and Gafarov F M 1999 \textit{Physica} \textbf{A274}
381
\bibitem{YulGafYulEme02} Ylmetyev R M, Gafarov F M, Yulmetyeva D G
and Emeljanova N A 2002 \textit{Physica} \textbf{A303} 427
\bibitem{YulEmeGaf04} Ylmetyev R M, Emeljanova N A and Gafarov F M 2004
\textit{Physica} \textbf{A341} 649
\bibitem{Zwa61} Zwanzig R 1961 \textit{Phys. Rev.} \textbf{124}
1338
\bibitem{Zwanzig65} Zwanzig R 1965 \textit{Annual review of physical
chemistry}\textbf{16}67
\end{thebibliography}
\end{document}